# Control of charge-spin interconversion in van der Waals heterostructures with chiral charge density waves

Zhendong Chi[1,*], Seungjun Lee[2,*], Haozhe Yang[1,*], Eoin Dolan[1], C. K. Safeer[1,3], Josep Ingla-Aynés[1], Franz Herling[1], Nerea Ontoso[1], Beatriz Martín-García[1,4], Marco Gobbi[4,5], Tony Low[2,6], Luis E. Hueso[1,4], Fèlix Casanova[1,4,†]

[1]CIC nanoGUNE BRTA, 20018 Donostia-San Sebastián, Basque Country, Spain

[2]Department of Electrical and Computer Engineering, University of Minnesota, Minneapolis, Minnesota 55455, USA

[3]Department of Physics, Clarendon Laboratory, University of Oxford, OX13PU, Oxford, United Kingdom

[4]IKERBASQUE, Basque Foundation for Science, 48009 Bilbao, Basque Country, Spain

[5]Centro de Física de Materiales (CSIC-EHU/UPV) and Materials Physics Center (MPC), 20018 Donostia-San Sebastián, Basque Country, Spain

[6]Department of Physics, University of Minnesota, Minneapolis, Minnesota 55455, USA

[*]These authors equally contributed to this work.

[†]E-mail: f.casanova@nanogune.eu

## ABSTRACT

A charge density wave (CDW) represents an exotic state in which electrons are arranged in a long range ordered pattern in low-dimensional materials. Although our understanding of the fundamental character of CDW has been enriched after extensive studies, its practical application remains limited. Here, we show an unprecedented demonstration of a tunable charge-spin interconversion (CSI) in graphene/1T-TaS$_2$ van der Waals heterostructures by manipulating the distinct CDW phases in 1T-TaS$_2$. Whereas CSI from spins polarized in all three directions are observed in the heterostructure when the CDW phase does not show commensurability, the output of one of the components disappears and the other two are enhanced when the CDW phase becomes commensurate. The experimental observation is supported by first-principles calculations, which evidence that chiral CDW multidomains in the heterostructure are at the origin of the switching of CSI. Our results uncover a new approach for on-demand CSI in low-dimensional systems, paving the way for advanced spin-orbitronic devices.

# 1. Introduction

Efficient and controllable charge-spin interconversion (CSI) obtained in materials with strong spin-orbit coupling is crucial for novel spin-orbitronic applications, such as non-volatile memory [1] and logic devices [2] [3]. Conventional CSI effects include the spin Hall effect (SHE) in bulk materials [4] and the Rashba-Edelstein effect (REE) in two-dimensional systems with broken inversion symmetry [5]. By these mechanisms, a spin current ($j_s$), in the former, or a spin density ($n_s$), in the latter, are generated by a charge current ($j_c$). Both effects obey reciprocity, leading to inverse effects with the same efficiency [6] [7]. In conventional materials with high symmetry, the directions of $j_c$ and spin polarization ($S$) are perpendicular. Although large CSI values have been demonstrated in various systems [8] [9] [10] [11], controlling the CSI efficiency to design advanced multifunctional devices remains challenging.

Van der Waals materials, which have high electronic/structural tunability, provide a versatile platform for investigating novel spin-orbitronic phenomena [12]. Graphene (Gr) is an excellent van der Waals material for transferring spin current [13], but its weak spin-orbit coupling limits the generation of CSI. To overcome this issue, Gr can be combined with transition metal dichalcogenides (TMDCs) to form van der Waals heterostructures [14] [15] in which the strong spin-orbit coupling in the TMDC is imprinted onto Gr by proximity effect. Large CSI in Gr/TMDC heterostructures has been predicted [16] [17] [18] [19] and subsequently confirmed in experiments [20] [21] [22] [23]. An unconventional CSI, with $j_c$ collinear to $S$, has been observed in some Gr/TMDC systems [24] [25] [26] [27] [28] and attributed to the broken mirror symmetries at the twisted interface [29] [30] [31] [32]. A remarkable characteristic of Gr/TMDC heterostructures is their tunable CSI efficiency by using a gate voltage to shift the Fermi level in Gr [21] [22]. However, controlling the CSI using the degrees of freedom in the TMDC counterpart has not yet been achieved.

The charge density wave (CDW), a long-range ordered state of electrons observed in low-dimensional materials including TMDC, exhibits a thermally activated modulation. Beyond its wide exploration for the fundamental connections with some correlated states of matter, such as unconventional superconductivity [33], its relationship with functional phenomena might be a promising direction. Here, we explore 1T-TaS$_2$, a TMDC featuring helical CDW phases with various degrees of commensurability [34, 35, 36], as an ideal platform for controlling CSI. By spin precession experiments, we confirm the CSI with the spins polarized in the three directions predicted in the twisted Gr/1T-TaS$_2$ heterostructure [32]. Furthermore, we observe the ON and OFF of the unconventional CSI component by tuning the commensurability of the CDW phase (Fig. 1a). Our first-principles calculations based on density functional theory reveal that the disappearance of the unconventional CSI arises from the presence of chiral CDW multidomains in the commensurate phase in the heterostructure, where the sign of the CSI is locked to chirality. The rich interplay between proximity, commensurability, and chirality of the CDW in the Gr/1T-TaS$_2$ van der Waals heterostructures opens the path to tailor a plethora of spin-based phenomena in low-dimensional systems.

# 2. Results and Discussion

## 2.1 Gr/1T-TaS$_2$ heterostructure CSI devices

1T-TaS$_2$ undergoes a reversible incommensurate CDW (ICCDW) to nearly commensurate CDW (NCCDW) phase transition and a NCCDW to commensurate CDW (CCDW) phase transition by modulating the temperature [34], displaying a resistance hysteresis loop during a temperature sweep (Fig.1b). In the CCDW phase, 1T-TaS$_2$ is filled with a star-of-David CDW lattice distortion from the normal state (Fig. 1a), exhibiting an insulating-like behavior. However, several tens of star-of-David CDW domains exist in the NCCDW phase while slight Ta-atom-position-shifts are observed in the ICCDW phase (see insets of Fig. 1b). The electrical transport in these two phases is thus limited by the normal metallic state [37]. This statement is also supported by a recent scanning tunnel microscope (STM) study, where the conductance across the domain walls of distinct insulating CCDW domains in 1T-TaS$_2$ is observed to be significantly higher than that within the CCDW domains [38]. Note that, although the band structures of 1T-TaS$_2$ in the ICCDW and NCCDW phases have been observed to show significant differences [39] [40], the case should be different in Gr/1T-TaS$_2$ heterostructures due to the charge transfer/carrier doping from Gr to the 1T-TaS$_2$ layer. The nature of the interfacial environment and transferred spin-orbit coupling from 1T-TaS$_2$ to Gr are contingent upon the CDW phase of 1T-TaS$_2$ [32]. Therefore, tunable CSI is expected in Gr/1T-TaS$_2$.

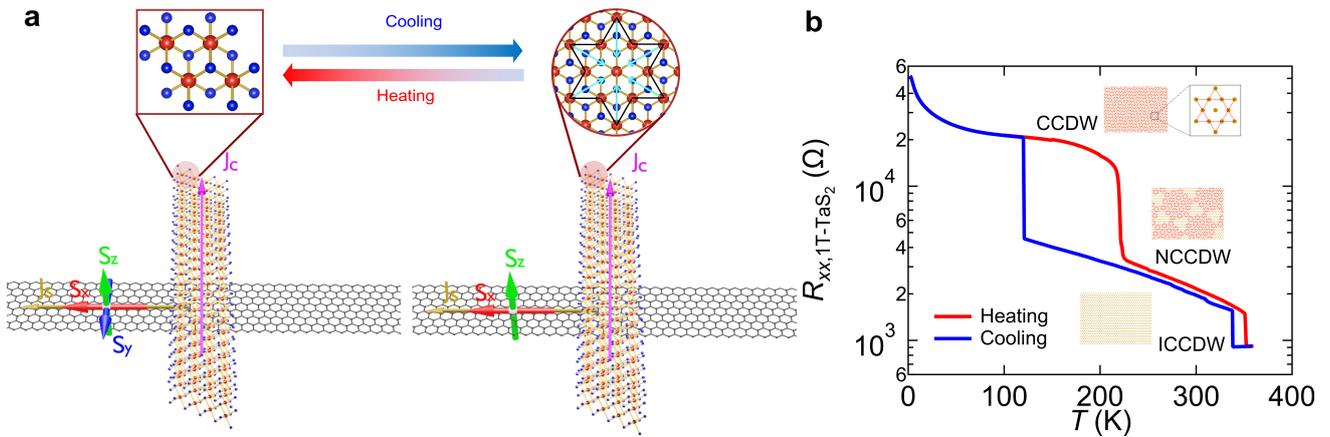

**Figure 1. Graphene/1T-TaS$_2$ crystal structure and the charge-spin interaction. a**, Schematic illustrations of charge-spin interconversion in graphene/1T-TaS$_2$ heterostructures in different 1T-TaS$_2$ phases. The spin polarization of the spin current ($j_s$, yellow arrow) along the $x$-, $y$-, and $z$-axes are labelled by red, blue, and green arrows. Top panel: top views of 1T-TaS$_2$ crystal structure in the normal (left) and CDW (right) states. The black lines in the sketch of the CDW state display the structure of the stars of David in 1T-TaS$_2$. The cyan arrows illustrate the distortion of the surrounding Ta atoms towards the central Ta atom. The middle arrows represent cooling (blue) and heating (red) cases. **b**, Temperature ($T$) dependent resistance of the 60-nm-thick 1T-TaS$_2$ nanocrystal ($R_{xx,1T-TaS_2}$) in Device 1 while cooling (blue) and heating (red). Insets: Formation of the stars of David in different CDW phases.

To thoroughly investigate the CSI in Gr/1T-TaS$_2$ heterostructures, we fabricated 3 different devices (see Methods and Supplementary Notes 1-3 for details on device fabrication and characterization). Figure 2a illustrates the measurement configuration of CSI. The device consists of a Gr/1T-TaS$_2$ Hall bar heterostructure, ferromagnetic (FM) electrodes, and non-magnetic (NM) electrodes. The FM electrodes are designed as elongated nanowires to obtain a magnetic easy axis along the *y*-axis. In our measurement, a dc charge current ($I_{dc}$) is applied across the 1T-TaS$_2$ layer. In the presence of CSI in the heterostructure, the charge current is converted into a spin current which diffuses in Gr along the *x*-axis while keeping its spin polarization ($S_x$, $S_y$, and $S_z$ shown in Fig. 2a). Such spin current generates a non-local voltage ($V_{NL}$) at the FM/Gr interface, which we detect between the FM and NM electrodes on Gr. The non-local CSI resistance is defined as $R_{NL} = V_{NL}/I_{dc}$. To disentangle the CSI components induced by different spin polarizations ($R_{CSI}^{S_x}$, $R_{CSI}^{S_y}$, and $R_{CSI}^{S_z}$), we measure $R_{NL}$ by sweeping an external magnetic field along the *x*- ($H_x$) and *z*-axes ($H_z$), inducing Hanle precession of the spin current [20] (See Methods for details). The magnetic properties of the FM electrodes and the spin transport characteristics of the pristine Gr and Gr/1T-TaS$_2$ heterostructure have been characterized individually in the same device using non-local spin valve and Hanle precession measurements (see Supplementary Note 5 for details) [41]. An electrical gate voltage ($V_G$) was applied to Gr to tune its Fermi level.

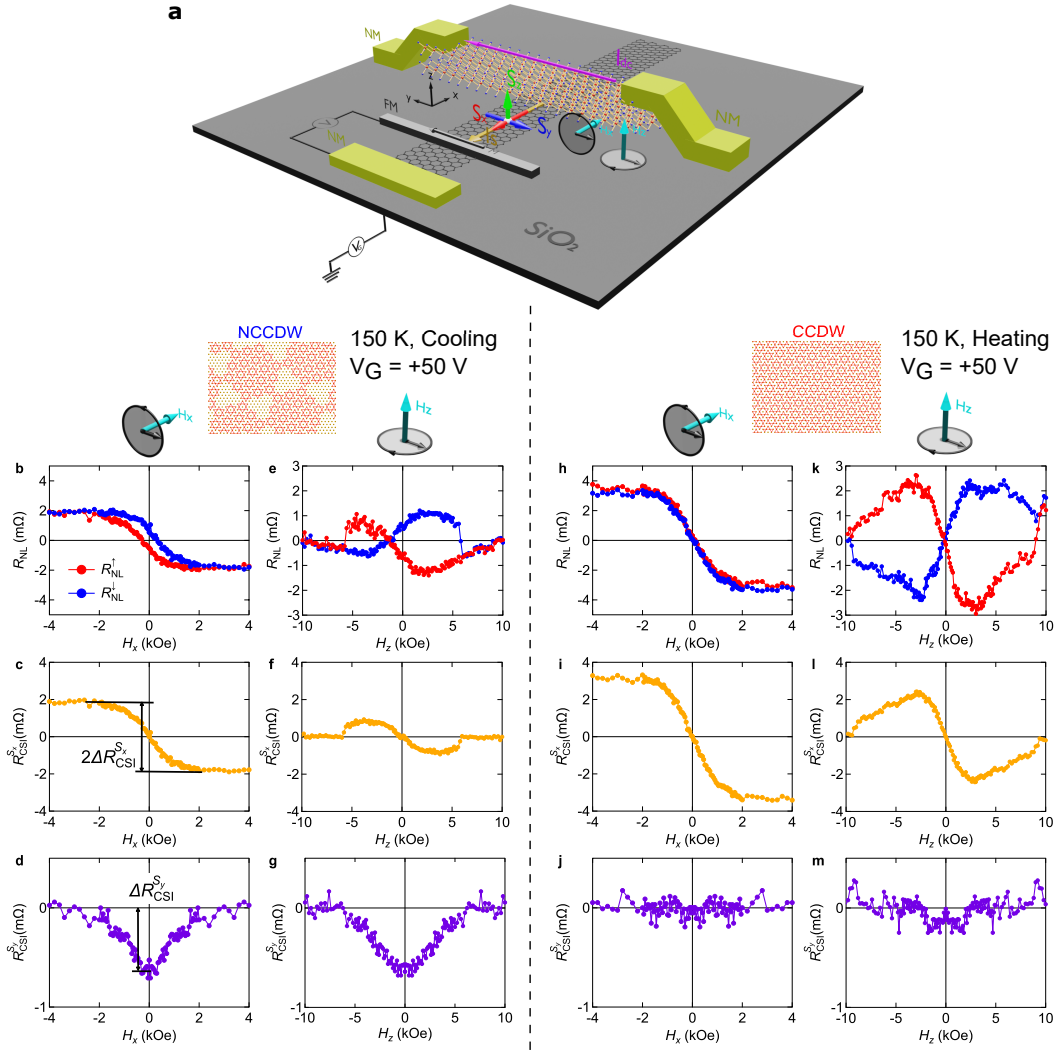

**Figure 2. CSI characterization of Gr/1T-TaS$_2$ heterostructure Device 1 with different 1T-TaS$_2$ CDW phases. a**, Schematic illustration of the CSI device. The device consists of a Gr/1T-TaS$_2$ heterostructure Hall bar cross junction, ferromagnetic (FM) nanowire electrodes (grey), and non-magnetic (NM) electrodes (light

green). The bottom-right sketches illustrate the Hanle precession of the spin current under magnetic fields applied along the x ($H_x$) and z ($H_z$) axes. The gate voltage ($V_G$) is applied through the SiO$_2$ layer of the substrate. **b**, Non-local CSI resistance ($R_{NL}$) measured as a function of $H_x$ with the NCCDW phase of 1T-TaS$_2$. The initial magnetization directions of the FM electrode are set along the +y ($R_{NL}^\uparrow$, red line) and −y directions ($R_{NL}^\downarrow$, blue line). **c**, $H_x$-dependent CSI component induced by spin current with spin polarization along the x-axis ($R_{CSI}^{S_x}$) extracted by averaging the two curves in b. The amplitude of $R_{CSI}^{S_x}$ is labeled as $\Delta R_{CSI}^{S_x}$. **d**, $H_x$-dependent CSI component induced by spin current with spin polarization along the y-axis ($R_{CSI}^{S_y}$) obtained by symmetrizing the half of the difference of the two curves in b. The amplitude of $R_{CSI}^{S_y}$ is labeled as $\Delta R_{CSI}^{S_y}$. **e**, $R_{NL}^\uparrow$ (red line) and $R_{NL}^\downarrow$ (blue line) measured as a function of $H_z$ with the NCCDW phase of 1T-TaS$_2$. The sudden jump of the two curves at $|H_z|$ ~5 kOe is attributed to a magnetization switching of the FM electrode, but does not affect the spin precession feature of the curves. **f**, $H_z$-dependent $R_{CSI}^{S_x}$ obtained by antisymmetrizing the half of the difference of the two curves in e. **g**, $H_z$-dependent $R_{CSI}^{S_y}$ extracted by symmetrizing the half of the difference of the two curves in e. **h**, $R_{NL}^\uparrow$ (red line) and $R_{NL}^\downarrow$ (blue line) measured as a function of $H_x$ with the CCDW phase of 1T-TaS$_2$. **i**, $H_x$-dependent $R_{CSI}^{S_x}$ acquired by averaging the two curves in h. **j**, $H_x$-dependent $R_{CSI}^{S_y}$ obtained by symmetrizing the half of the difference of the two curves in h. **k**, $R_{NL}^\uparrow$ (red line) and $R_{NL}^\downarrow$ (blue line) measured as a function of $H_z$ with the CCDW phase of 1T-TaS$_2$. **l**, $H_z$-dependent $R_{CSI}^{S_x}$ acquired by antisymmetrizing the half of the difference of the two curves in k. **m**, $H_z$-dependent $R_{CSI}^{S_y}$ obtained by symmetrizing the half of the difference of the two curves in k. A constant baseline of 1.9, 2.9, 6.8, and 8.6 mΩ has been subtracted from the curves in b, e, h, and k, respectively. All data was collected at 150 K with a gate voltage $V_G$ = +50 V.

2.2 CDW phase dependence of CSI

Figure 2 illustrates the dependence of the CDW phase on the CSI characterization in Device 1 (bilayer Gr/60-nm-thick 1T-TaS$_2$) at $V_G$ = +50 V. All the measurements were performed at 150 K, where either the NCCDW or the CCDW phase of 1T-TaS$_2$ can be stabilized by cooling or heating (Fig. 1b), while the Gr properties remain constant. Figure 2b displays the $H_x$-dependent $R_{NL}$ at the NCCDW phase. The red ($R_{NL}^\uparrow$) and blue ($R_{NL}^\downarrow$) curves were measured with the magnetization ***M*** of the FM electrode (Fig. 2a) initially set along the +y and −y directions. To extract the different CSI components, we define $R_{NL}^{avg}(H_x) = (R_{NL}^\uparrow + R_{NL}^\downarrow)/2$ and $R_{NL}^{diff}(H_x) = (R_{NL}^\uparrow - R_{NL}^\downarrow)/2$. This average corresponds to the CSI component of the spins polarized along the x-axis ($R_{CSI}^{S_x}(H_x) \equiv R_{NL}^{avg}(H_x)$) and the difference includes both the CSI components of the spins polarized along the y- ($R_{CSI}^{S_y}$) and z-axes ($R_{CSI}^{S_z}$). $R_{CSI}^{S_y}$ and $R_{CSI}^{S_z}$ can be further distinguished by symmetrizing and antisymmetrizing $R_{NL}^{diff}(H_x)$, respectively (see Methods for details).

The extracted $R_{CSI}^{S_x}$ is presented as a function of $H_x$ in Fig. 2c, revealing an S-shaped curve. The value of $R_{CSI}^{S_x}$ increases with increasing $|H_x|$ and reaches a maximum of 1.9 mΩ at $|H_x|$~2 kOe, where ***M*** is fully saturated along the x-axis. Figure 2d illustrates the $H_x$-dependence of $R_{CSI}^{S_y}$, which has a maximum of 0.7 mΩ when $H_x$ is zero and decreases as $H_x$ increases until it vanishes at $|H_x|$ ~ 2 kOe. This CSI component is confirmed by measuring $R_{NL}$ as a function of $H_y$, which shows a squared hysteresis loop

(See Supplementary Note 7 for details). $R_{CSI}^{S_z}$ is not shown, as it is below the noise level. Note that the S-shaped curve of $R_{CSI}^{S_x}$ with $H_x$ could also be a spurious effect originating from the ordinary Hall effect in Gr induced by the stray field of the FM electrode [42]. To confirm the spin transport origin of the signal, i.e., CSI, $R_{NL}$ was measured by sweeping $H_z$ to obtain the spin precession in the $x$–$y$ plane. Figure 2e displays $R_{NL}^{\uparrow}$ and $R_{NL}^{\downarrow}$ as a function of $H_z$. In this case, $R_{CSI}^{S_x}$ and $R_{CSI}^{S_y}$ are extracted by defining $R_{NL}^{diff}(H_z) = (R_{NL}^{\uparrow} - R_{NL}^{\downarrow})/2$ and taking the antisymmetric and symmetric part, respectively (see Methods). Figure 2f shows that $R_{CSI}^{S_x}$ exhibits a clear antisymmetric Hanle precession feature vs. $H_z$, reaching a maximum of 0.9 mΩ at $H_z \sim 3$ kOe. $R_{CSI}^{S_y}$ is plotted as a function of $H_z$ in Fig. 2g, showing a maximum of 0.7 mΩ when $H_z$ is zero and decreasing with increasing $H_z$. The maxima of $R_{CSI}^{S_y}$ in both $H_x$- and $H_z$-dependent measurements are in good agreement. These results provide unequivocal proof that the CSI signals in Figs. 2b and 2e are induced by the CSI components of spin currents polarized along the $x$- and $y$-axes. We determine the amplitude of $R_{CSI}^{S_x}$ and $R_{CSI}^{S_y}$ using, respectively, the step height in Fig. 2c ($\Delta R_{CSI}^{S_x} \equiv \left(R_{CSI}^{S_x}(H_x = 2 \text{ kOe}) - R_{CSI}^{S_x}(H_x = -2 \text{ kOe})\right)/2$) and the maximum value in Fig. 2d ($\Delta R_{CSI}^{S_y} \equiv \left|R_{CSI}^{S_y}(H_x = 0)\right|$).

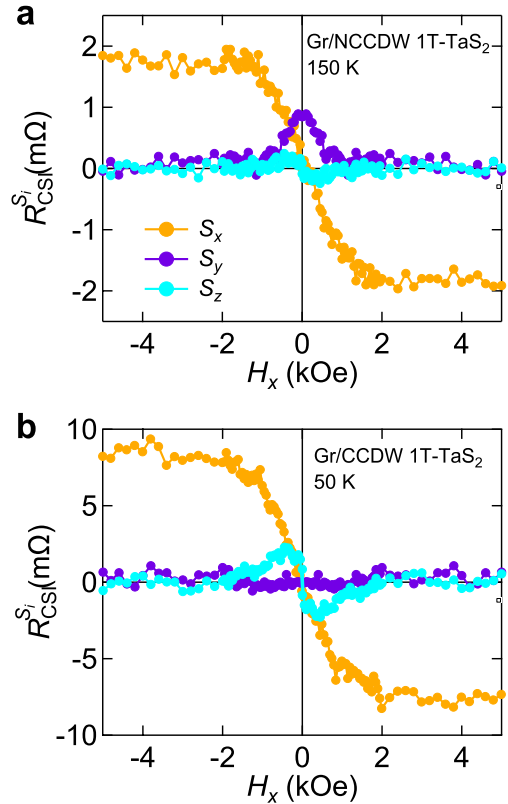

Figure 3. Spin-polarization direction dependent CSI in Device 2. Different components of the CSI ($R_{CSI}^{S_i}$) extracted from the non-local CSI resistance measured as a function of $H_x$ in Device 2. $R_{CSI}^{S_i}$ in Gr/NCCDW 1T-TaS$_2$ heterostructure at 150 K and in Gr/CCDW 1T-TaS$_2$ heterostructure at 50 K are displayed in **a** and **b**, respectively. $R_{CSI}^{S_i}$ originating from spin currents with spin polarization along the $x$ ($R_{CSI}^{S_x}$), $y$ ($R_{CSI}^{S_y}$), and $z$ ($R_{CSI}^{S_z}$)

axes are plotted as orange, purple, and cyan curves, respectively. All the measurements were performed without applying a gate voltage. See Supplementary Note 6 for the raw data.

Similar measurements were conducted after driving the NCCDW-CCDW phase transition in 1T-TaS$_2$ and stabilizing the CCDW phase at 150 K. The resulting $R_{NL}$ and the corresponding $R_{CSI}^{S_x}$ and $R_{CSI}^{S_y}$ are plotted as a function of $H_x$ (Figs. 2h-2j) and $H_z$ (Figs. 2k-2m). The $S$-shaped curve (antisymmetric Hanle precession) of $R_{CSI}^{S_x}$ as a function of $H_x$ ($H_z$) indicates that the CSI component induced by $x$-polarized spin current persists in the CCDW phase. $\Delta R_{CSI}^{S_x}$ increases by approximately a factor of two after the phase transition. Interestingly, regardless of whether the magnetic field is applied along the $x$- or $z$-axis, $R_{CSI}^{S_y}$ is nearly zero. This result evidences that the unconventional CSI is switched off in our heterostructure when 1T-TaS$_2$ is in the CCDW phase.

To investigate whether the disappearance of the unconventional CSI in Gr/1T-TaS$_2$ heterostructure during the phase transition is reproducible, we repeated the experiment in another device (Device 2, monolayer Gr/100-nm-thick 1T-TaS$_2$). $R_{CSI}^{S_x}$, $R_{CSI}^{S_y}$, and $R_{CSI}^{S_z}$ obtained with the NCCDW and CCDW phases are plotted as a function of $H_x$ in Figs. 3a and 3b, respectively. We find that $R_{CSI}^{S_y}$ disappears again after the phase transition while $R_{CSI}^{S_x}$ exists regardless of the CDW phase, confirming the previous observation.

Interestingly, although $R_{CSI}^{S_z}$ in Device 1 is lower than the noise level, we clearly observe $R_{CSI}^{S_z}$ in Device 2 when 1T-TaS$_2$ is in either the NCCDW or the CCDW phase, as shown by the clear antisymmetric Hanle precession cyan curves in Figs. 3a and 3b, respectively. This CSI component is also detected in another device (Device 3, few-layer Gr/22-nm-thick 1T-TaS$_2$), indicating that the number of Gr layers is not the reason for observing $R_{CSI}^{S_z}$. We note that the spin injection efficiency of the ferromagnet, estimated from non-local spin valve measurements, is much lower in Device 1 than in Devices 2 and 3 (See Supplementary Note 5). Compared to the relatively large $\Delta R_{CSI}^{S_x}$ and $\Delta R_{CSI}^{S_y}$, especially when 1T-TaS$_2$ is in the NCCDW phase (Fig. 3a), the lower spin injection efficiency in Device 1 prevents the detection of the small $R_{CSI}^{S_z}$ component.

We also investigated CSI in Device 3 when 1T-TaS$_2$ is in the ICCDW phase (See Supplementary Note 8 for details). All three CSI components are observed at 360 K. As discussed above, the transport in 1T-TaS$_2$ is principally governed by the normal state, both in the NCCDW and ICCDW phases. Conversely, in the CCDW phase, the CDW state predominates, thereby instigating insulating-like behavior. These results suggest that the phase-dependent switching of the unconventional CSI ($R_{CSI}^{S_y}$) is linked to the change in the transport regime (that is, commensurability) of 1T-TaS$_2$.

2.3 Temperature and gate voltage dependencies of CSI

After confirming an omnidirectional CSI and a phase-transition-driven switchable unconventional CSI, we studied the temperature ($T$) and gate voltage ($V_G$) dependencies of CSI. $\Delta R_{CSI}^{S_x}$ and $\Delta R_{CSI}^{S_y}$ extracted from Device 1 are plotted as a function of $T$ in Fig. 4a. We observe that both $\Delta R_{CSI}^{S_x}$ and $\Delta R_{CSI}^{S_y}$ increase as $T$ decreases, regardless of the phase of 1T-TaS$_2$. Furthermore, $\Delta R_{CSI}^{S_x}$ exhibits a phase-related hysteresis loop with $T$, which is consistent with the resistance hysteresis loop of 1T-TaS$_2$ (Fig. 1b). This agreement indicates that the resistance of 1T-TaS$_2$ influences the CSI output in Gr/1T-TaS$_2$ heterostructures.

Figure 4b illustrates $\Delta R_{CSI}^{S_x}$ and $\Delta R_{CSI}^{S_y}$ extracted from Device 1 at 150 K as a function of $V_G$. All the CSI components follow a similar trend with respect to $V_G$: their magnitudes reach a minimum when $V_G = -40$ V, i.e., the charge neutrality point of the Gr/1T-TaS$_2$ heterostructure (Fig. 4c), and increase as $V_G$ increases from $-40$ V. The values reach saturation after $V_G$ is increased up to $+40$ V. This behavior differs from that observed in the resistance of Gr/1T-TaS$_2$ heterostructure (Fig. 4c). To understand the observed $V_G$ dependence, we measured the charge and spin transport properties of Device 1 at 150 K with different $V_G$. The resistance of 1T-TaS$_2$ exhibits negligible changes with $V_G$ (Fig. S4b and S4c in Supplementary Note 4). In contrast, the spin injection efficiency, given by the non-local spin valve signal shown in Fig. 4d (see details on the measurement in Supplementary Note 5), shows a similar dependence on $V_G$ as the CSI output. Therefore, we suggest that the tuning of the CSI output by the gate voltage can be primarily attributed to the gate tuning of the spin injection efficiency at the Gr/FM electrode interface. Similar $T$ and $V_G$ dependencies of CSI are obtained in Device 3 (see Supplementary Note 8).

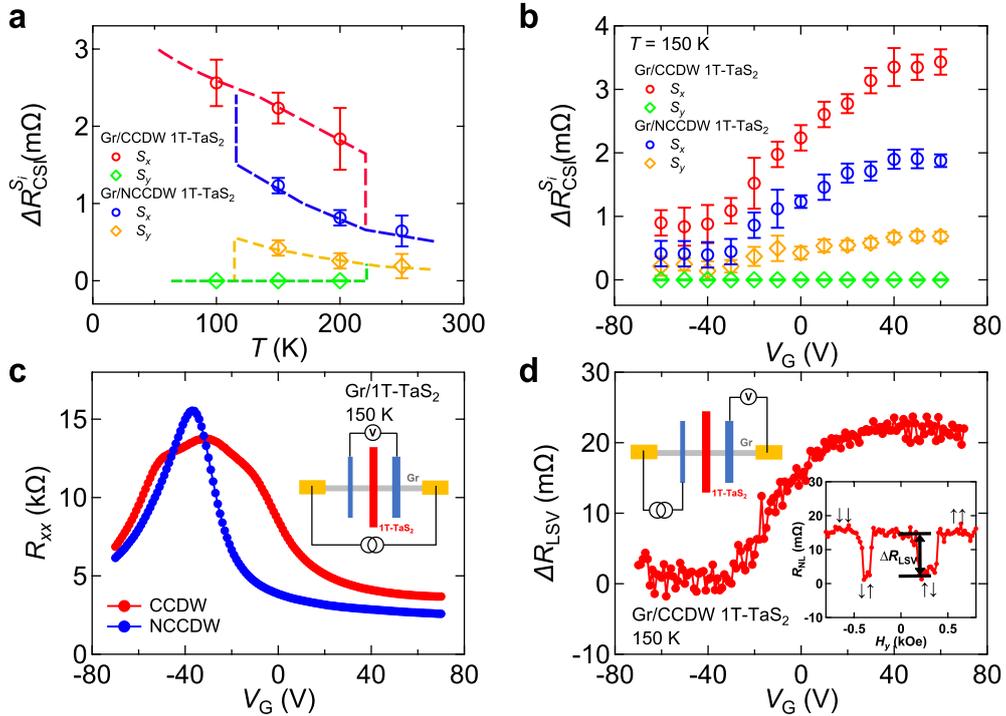

**Figure 4. Temperature and gate voltage dependencies of CSI in Device 1. a**, Temperature ($T$) dependence of $\Delta R_{CSI}^{S_i}$ measured without adding gate voltage. $\Delta R_{CSI}^{S_x}$ in the heterostructure in the NCCDW (CCDW) phase of 1T-TaS$_2$ are displayed by red (blue) circles. $\Delta R_{CSI}^{S_y}$ in the heterostructure in the NCCDW phase of 1T-TaS$_2$ is shown by orange diamonds. The vanished $\Delta R_{CSI}^{S_y}$ in the heterostructure in the CCDW phase of 1T-TaS2 is represented by the green diamonds, with all the values equal to zero. The dashed curves are a guide to the eye for the hysteresis loops observed in $\Delta R_{CSI}^{S_i}$. **b**, $V_G$ dependence of $\Delta R_{CSI}^{S_i}$ measured at $T = 150$ K. **c**, $V_G$ dependence of the longitudinal resistance of the Gr/1T-TaS$_2$ heterostructure ($R_{xx}$) when 1T-TaS$_2$ is at the CCDW (red circles) and NCCDW (blue circles) phases. Inset: Sketch of the measurement configuration. **d**, $V_G$ dependence of the non-local spin valve signal ($\Delta R_{LSV}$) measured across the Gr/CCDW 1T-TaS$_2$ heterostructure at 150 K. Insets: Sketch of the measurement configuration (left) and the non-local resistance of the spin valve ($R_{NL}$) measured with $V_G = 0\ V$ (right). The definition of $\Delta R_{LSV}$ is labelled. Error bars are calculated using the standard deviation associated with the statistical average of $R_{CSI}^{S_i}$ in both the positive and negative magnetic field ranges.

2.4 The origin of the CSI

The observed CSI in the Gr/1T-TaS$_2$ heterostructure can have different origins: the SHE inherent to the bulk 1T-TaS$_2$, the interfacial SHE arising from the spin-orbit proximity, and the REE generated by the broken inversion symmetry at the interface. To gain insight into the underlying mechanisms of tunable CSI driven by the commensurability of the CDW phase, we perform first-principles calculations based on density functional theory (see Methods) to analyze the CSI of bulk 1T-TaS$_2$ and of the Gr/1T-TaS$_2$ van der Waals heterostructure in the normal and the CCDW states of 1T-TaS$_2$.

We investigate first the SHE of bulk 1T-TaS$_2$ by calculating the spin Hall conductivity tensor $\sigma_{ij}^k$, where $i$, $j$, and $k$ represent the directions of $j_s$, $j_c$, and $S$, respectively. In our measurement, the charge current is along the $y$-axis, whereas the spin current direction in the bulk 1T-TaS$_2$ can be either along the $x$-axis ($j_{s,x}$) or along the $z$-axis ($j_{s,z}$) due to the absorption from Gr, as seen in metallic systems [41]. Table 1 summarizes the calculated $\sigma_{ij}^k$ values of possible elements in our measurement configuration, determined by the spin Berry curvature component $\Omega_{ij}^k$ (see Supplementary Note 9 for details). On the one hand, for $j_{s,x}$, $\sigma_{xy}^z$ is much larger than $\sigma_{xy}^x$ and $\sigma_{xy}^y$ in the normal state. These values suggest the $R_{CSI}^{S_z}$ would be the largest CSI component detected in the NCCDW or ICCDW phases of 1T-TaS$_2$. However, this prediction contradicts our experimental results, leading us to exclude the possibility of CSI originating from the SHE of bulk 1T-TaS$_2$ induced by $j_{s,x}$. On the other hand, for $j_{s,z}$, $\sigma_{zy}^y$ and $\sigma_{zy}^z$ are strictly zero (due to symmetry constrain [43]) or negligible for the normal and CCDW states, respectively, indicating that the conventional SHE ($\sigma_{zy}^x$) in bulk 1T-TaS$_2$ is the only bulk 1T-TaS$_2$-originated component that can contribute to the observed $R_{CSI}^{S_x}$.

**Table 1. Spin Hall conductivity in bulk 1T-TaS$_2$ and the graphene/1T-TaS$_2$ heterostructure calculated in the CCDW and normal phases of 1T-TaS$_2$.** The spin Hall conductivity is represented by $\sigma_{ij}^k$, where $i$, $j$, and $k$ represent the spin current, charge current, and spin polarization directions respectively. Since $\sigma_{ij}^k$ is very sensitive to the crystallographic orientation, each number indicates the maximum values of $\sigma_{ij}^k$ for any possible crystal orientations. All units are in $((\hbar/e)\ S/cm)$.

| | | $\sigma_{xy}^x$ | $\sigma_{xy}^y$ | $\sigma_{xy}^z$ | $\sigma_{zy}^x$ | $\sigma_{zy}^y$ | $\sigma_{zy}^z$ |
|---|---|---|---|---|---|---|---|
| 1T-TaS$_2$ | normal | 6.5 | 6.5 | 59.6 | 22.0 | - | - |
| | CDW | 9.0 | 8.0 | 24.0 | 15.0 | 3.0 | 1.0 |
| Gr/1T-TaS$_2$ | normal | 6.0 | 6.0 | 53.0 | - | - | - |
| | CDW | 15.0 | 20.0 | 18.0 | - | - | - |

Next, we investigate the interfacial SHE of the Gr/1T-TaS$_2$ heterostructure with a twisted angle of 13.9°. The calculated $\sigma_{ij}^k$ values are summarized in Table 1 (see Supplementary Note 9 for details). By nature of the interfacial effect, the direction of the spin current is constrained to the x-axis. In the calculated spin Berry curvature, it becomes evident that the Dirac band of Gr mostly contributes to the integration when the 1T-TaS2 is in the CCDW phase. Conversely, the contribution from 1T-TaS$_2$ bands predominates in the Gr/normal state 1T-TaS$_2$ heterostructure (See Supplementary Note 9 for details). Such feature implies that the source of the interfacial SHE shifts from the 1T-TaS$_2$ side to the Gr side following the transition from the NCCDW to the CCDW phase. In the normal state, the conventional component $\sigma_{xy}^z$ shows a much larger value compared to the unconventional components $\sigma_{xy}^x$ and $\sigma_{xy}^y$. However, its contribution to the CSI output should be limited because most of the current flows in the bulk 1T-TaS$_2$. In the CCDW state, $\sigma_{xy}^z$ is smaller than that in the normal state. In this case, the current redistributes and passes predominantly through the Gr side due to the enhanced resistance of CCDW 1T-TaS$_2$, indicating that the CSI output should increase. These properties are consistent with $\Delta R_{CSI}^{S_z}$ observed in Device 2 (Fig. 3), suggesting that the CSI induced by $S_z$-spin current is due to the conventional interfacial SHE. Note that unconventional $\sigma_{xy}^x$ and $\sigma_{xy}^y$ are comparable to $\sigma_{xy}^z$ in the CCDW state, indicating that they can contribute to $R_{CSI}^{S_x}$ and $R_{CSI}^{S_y}$, respectively. However, the finite $\sigma_{xy}^y$ is inconsistent with the vanishing $R_{CSI}^{S_y}$ in our experiment.

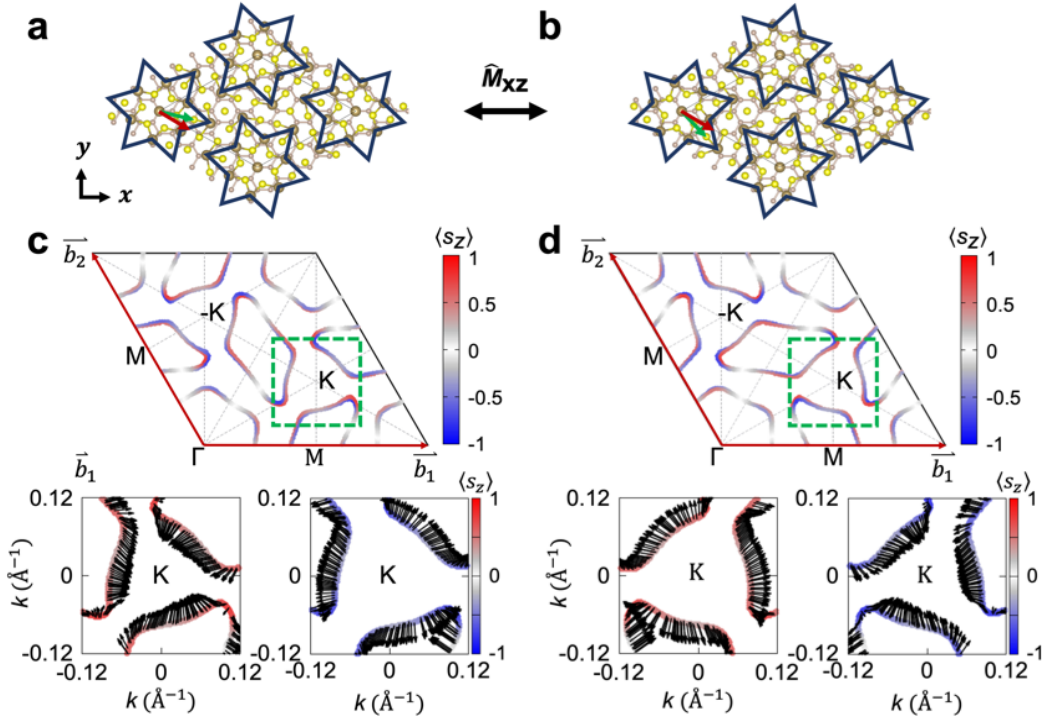

**Figure 5. First-principles calculations for Rashba spin texture in Gr/1T-TaS$_2$ heterostructures. a, b,** Two possible chiral states of Gr/CCDW 1T-TaS$_2$ heterostructures with opposite handedness. The green and red arrows indicate unit vector of normal and CCDW 1T-TaS$_2$, respectively, which have either +13.9° or −13.9° rotation angle. They are mirror symmetric corresponding to the $\hat{M}_{xz}$ mirror operator. **c, d,** Handedness-dependent spin-resolved Fermi surfaces and Rashba spin textures of proximitized Dirac bands. Near the K point

(highlighted by a green box), red and blue colors show spin splitting in the Dirac bands due to the Rashba spin-orbit coupling. The lower two panels show in-plane spin textures of inner (left) and outer (right) Rashba bands. Only the radial spin components are opposite in the chiral structures with opposite handedness.

Finally, we consider the REE as the source of our observed CSI. On the one hand, the conventional REE can contribute to $R_{\text{CSI}}^{S_x}$ in our setup. On the other hand, recent studies have revealed that $R_{\text{CSI}}^{S_y}$ in Gr/TMDC van der Waals heterostructures with a twist angle [26] [27] is related to an unconventional REE (UREE) induced by radial spin textures in the Rashba bands due to the lack of a mirror plane [29] [30] [31] [32]. Indeed, we observed a radial spin texture in both Gr/CCDW 1T-TaS$_2$ and Gr/normal 1T-TaS$_2$ heterostructures (Fig. 5 and Fig. S17 in Supplementary Note 9) in our calculations. Since UREE can contribute to $R_{\text{CSI}}^{S_y}$, it is reasonable to expect a non-zero $R_{\text{CSI}}^{S_y}$ in both normal and CCDW 1T-TaS$_2$ states. This result, as in the case of the finite interfacial $\sigma_{xy}^y$, also contradicts our experimental observation.

To reconcile the theory with experimental observations, we examine the role of chirality in the Gr/CCDW 1T-TaS$_2$ heterostructures, as described in Figs. 5a and 5b. It is well known that the lattice vectors of CCDW 1T-TaS$_2$ are rotated by either +13.9° or −13.9° from normal 1T-TaS$_2$, resulting in two helical states with opposite handedness [40]. These helical states are mirror-symmetric with respect to any mirror plane perpendicular to the TaS$_2$ layer and can be simultaneously observed in a single 1T-TaS$_2$ flake [40]. Note that this helical domain is intrinsic and is only described by mirroring, not by rotation, and is distinct from a trivial polycrystalline domain. In the presence of Gr/CCDW 1T-TaS$_2$ interface, the inversion symmetry is broken. This leads to the transformation of the helicity of 1T-TaS$_2$ into the chirality of the resulting heterostructures. In general, arbitrary twist angles between Gr and 1T-TaS$_2$ may break the mirror relation between two chiral heterostructures. However, to better grasp the underlying physics, we focused on simple examples that exhibit the mirror relation between two chiral heterostructures, ($\widehat{M}_{xz}$, in our case) and investigated the role of chiral domains.

We present the spin-resolved Fermi surfaces of both chiral heterostructures in Figs. 5c and 5d. Near the K point (indicated by a green box), we observed a Gr Dirac band with distinct spin splitting, resulting in inner (red) and outer (blue) Rashba bands. The in-plane spin textures of both Rashba bands are shown in the lower two panels. Our results reveal opposite radial spin textures for different handedness, while the conventional Rashba component remains the same in both chiral states, which means that the radial spin is locked to the chirality [44] [45][42][43]. (See Figs. S18 and S19 for more details) This is because the $\widehat{M}_{xz}$ operation flips $S_x$, $S_z$, and $k_y$, but not $S_y$ and $k_x$. Here $k_x$ ($k_y$) represents the momentum along the $x(y)$-axis. These results imply that conventional REE is preserved whereas UREE should disappear when both chiral CDW structures coexist. The effect of chiral domain on the interfacial SHE can also be deduced from the quantity $S_{ij}^k = v_i S_k v_j$, where $v$ represents the velocity of electrons [43]. This quantity has the same symmetry as $\Omega_{ij}^k$. Similar to UREE, $\widehat{M}_{xz}\Omega_{xy}^y = v_x S_y(-v_y) = -\Omega_{xy}^y$, indicating that the total interfacial $\sigma_{xy}^y$ is also canceled out if the two handedness coexist. We have conducted additional calculations on the Gr/CCDW 1T-TaS$_2$ heterostructure with a twisted angle of 5.2°. The results show that the radial spin textures are consistently opposite in the two structures with reversed handedness (Fig. S20), evidencing the robustness of the observed cancellation.

Note that the CDW domains in the NCCDW phase of 1T-TaS$_2$ exhibit an average size of ~10 nm at 200 K [46] [47]. The chiral domain expands and merges during the NCCDW-CCDW phase transition, which can increase its size in the CCDW phase. However, this scale is much smaller than our micrometer-level devices. Although nanometer- and millimeter-size chiral domains have been reported in recent studies [36, 48, 49], the size could also be restricted by the appearance of impurities [40]. Therefore, we deduce that the contributions from UREE and interfacial $\sigma_{xy}^y$ are canceled out due to the mixed chiral domains with opposite handedness, i.e., the vanishing $R_{\text{CSI}}^{S_y}$ in our experiments, whereas the conventional REE and interfacial $\sigma_{xy}^x$ contribute to the observed $R_{\text{CSI}}^{S_x}$. Hence, it can be concluded that the intrinsic chirality of the CCDW phase governs the evolution of both UREE and the interfacial unconventional SHE. Alternative hypothesis, for instance a non-uniform current distribution along the *x*-axis arising from the high-conductivity normal-phase channels in between the commensurate star-of-David domains of the NCCDW phase, could potentially generate the $R_{\text{CSI}}^{S_y}$ component by the conventional SHE. However, we maintain a net zero current along the x-axis during the measurements, ruling out this scenario as the origin of the unconventional CSI.

In combination with the control of the commensurability [34] [35] and the helicity [36] of the CDW phase in 1T-TaS$_2$ achieved via electrical stimulation, our observation provides new opportunities for designing functional electric-field-control-of spintronic devices with a degree of freedom from the material itself. For instance, the tunable CSI could be exploited in spin-orbit torque memory devices, where only specific unconventional CSI components contribute to the deterministic switching of perpendicular magnetization. This process has been demonstrated in low-symmetry materials [50] [51] [52] [53], where the underlying symmetry of CSI phenomena, and the allowed charge-spin response tensor, is governed by the crystal symmetry [43] [54]. The present 1T-TaS$_2$, with tunable CDW between high and low symmetry crystalline phases, simplifies such geometric obstacle and offers an enticing platform for realizing such on-demand switchable CSI.

## 3. Conclusion

In summary, we report the first demonstration of tunable CSI in Gr/1T-TaS$_2$ van der Waals heterostructures exploiting the phase transition of the CDW in the TMDC. We observe omnidirectional CSI (i.e., with spins oriented in all three directions) in the heterostructures when the CDW in 1T-TaS$_2$ does not show commensurability. Whereas the CSI components induced by $S_x$- and $S_z$-polarized spin currents are enhanced after a phase transition to a commensurate state, the one induced by $S_y$-polarized spin currents is switched off. By comparing our experimental results with first-principles calculations, we conclude that the disappearance of this CSI component is due to the appearance of chiral CDW multidomains. Our findings provide insights into the interplay between proximity, commensurability and chirality of the CDW in the Gr/1T-TaS$_2$ system and establish it as a platform for low-dimensional functional spin-orbitronic devices.

# 4. Experimental Section

Device fabrication

The graphene/1T-TaS$_2$ heterostructures were assembled by the dry viscoelastic transfer technique. Elongated graphene flakes were first exfoliated from highly oriented pyrolytic graphite (HQ Graphene) and deposited on highly *n*-type doped Si substrates with a 300-nm-thick SiO$_2$ layer. The monolayer and bilayer graphene flakes were distinguished based on the contrast using an optical microscope and confirmed by Raman spectroscopy. The 1T-TaS$_2$ flakes were exfoliated from a single crystal provided by 2D Semiconductors (Device 1) and by HQ Graphene (Devices 2 and 3) onto a piece of polydimethylsiloxane (PDMS). After identifying strips with suitable width and thickness (confirmed later by an atomic force microscopy) using an optical microscope, the 1T-TaS$_2$ flakes were then transferred from the PDMS stamp to the SiO$_2$/Si substrate, where they formed Hall bar cross junctions with the elongated graphene flakes. The exfoliation of 1T-TaS$_2$ flakes and transfer processes were carried out in a glovebox with an Ar atmosphere. Non-magnetic Pd (5 nm)/Au (45 nm) electrodes were used to contact the graphene and 1T-TaS$_2$ and fabricated using the standard processes of e-beam lithography, evaporation, and lift-off. TiO$_x$/Co/Au ferromagnetic (FM) electrodes were deposited on the graphene flake to inject/detect the spin current and fabricated using the same processes as Pd/Au electrodes. During the evaporation process, ~0.3 nm of Ti was first deposited, followed by a 10-min ambient oxidization process that formed the TiO$_x$ layer. Then, 35 nm of Co and 15 nm of Au were evaporated. The 15-nm-thick Au layer was a capping layer to protect the Co layer from oxidization. The width of FM electrodes was either 150 or 350 nm to give different shape anisotropies.

Electrical measurements

Local and non-local magnetotransport measurements were performed in a physical property measurement system (PPMS) by Quantum Design, where temperature was varied and magnetic fields were applied in different directions. The measurements were carried out using a dc reversal technique with a Keithley 2182 nanovoltmeter and a 6221 current source. A back-gate voltage was applied to the highly *n*-type doped Si substrate using a Keithley 2636B source meter unit to adjust the Fermi level of the graphene.

Separation of the different CSI components

In our CSI measurements, the spin polarization ($S$) of the spin current ($j_s$) diffusing in graphene is controllable by a magnetic field. By applying an external magnetic field along the *x*-axis ($H_x$) or *z*-axis ($H_z$), which respectively correspond to the in-plane and out-of-plane hard axes of the FM electrode, the spin in the spin current undergoes precession in the *y-z* or *x-y* plane. These precessions give additional spin-polarization components in the spin current. The magnetization *M* of the FM electrode is also pulled by the external magnetic field. The amplitude of different CSI components ($\Delta R_{\text{CSI}}^{S_i}$) is thus proportional to the projection of *M* along the corresponding spin polarization direction ($S_i$) of the spin current. To distinguish the components induced by a spin current with different polarizations, the non-local CSI resistance $R_{\text{NL}}$ is measured as a function of $H_x$ or $H_z$.

By scanning $H_x$, the $R_{NL}$ component induced by $S_x$ ($R_{CSI}^{S_x}$) is zero when $H_x = 0$ because the initial $M$ is perpendicular to $S_x$. The value of $R_{CSI}^{S_x}$ increases as $H_x$ increases and finally saturates when $M$ is fully aligned along the $x$-axis. This yields an $S$-shaped curve similar to the conventional SHE in metals [41] or the REE in 2D systems [21]. $R_{CSI}^{S_x}$ is unaffected by the initial direction of $M$ (+$y$ or –$y$) as its value is only determined by the projection of $M$ along the $x$-axis. On the other hand, because the spin current precesses within the $y$-$z$ plane under a small $H_x$, the $R_{NL}$ components originating from $S_y$ ($R_{CSI}^{S_y}$) and $S_z$ ($R_{CSI}^{S_z}$) show a spin precession feature as a function of $H_x$. Considering the initial $M$ along the $y$-axis, the precession of $R_{CSI}^{S_y}$ and $R_{CSI}^{S_z}$ should show symmetric and antisymmetric line shapes, respectively, as reported in other Gr/TMDC heterostructures [20] [21] [22] [24] [26] [27]. The signs of $R_{CSI}^{S_y}$ and $R_{CSI}^{S_z}$ reverse after switching the initial direction of $M$ because the initial projections of $M$ onto the $y$-axis are opposite. Both $R_{CSI}^{S_y}$ and $R_{CSI}^{S_z}$ will decrease to zero at high $H_x$ due to the $x$-axis pulling of $M$ and the spin dephasing. Based on these properties, we can separate $R_{CSI}^{S_x}$, $R_{CSI}^{S_y}$, and $R_{CSI}^{S_z}$ by measuring $R_{NL}$ with the initial direction of $M$ set along the +$y$ ($R_{NL}^\uparrow$) and –$y$ ($R_{NL}^\downarrow$) directions. $R_{CSI}^{S_x}$ can be extracted by taking their average ($R_{NL}^{avg}(H_x) = (R_{NL}^\uparrow + R_{NL}^\downarrow)/2$), since the sign of $R_{NL}^{S_x}$ is the same in both cases. The contribution of $R_{CSI}^{S_x}$ can be removed by taking half of the difference ($R_{NL}^{diff}(H_x) = (R_{NL}^\uparrow - R_{NL}^\downarrow)/2$), where the $R_{CSI}^{S_y}$ and $R_{CSI}^{S_z}$ contributions remain because their signs are opposite after reversing the initial direction of $M$. $R_{CSI}^{S_y}$ and $R_{CSI}^{S_z}$ can be further separated by symmetrizing and antisymmetrizing $R_{NL}^{diff}(H_x)$, respectively, due to their different precession features.

The line shapes of $R_{NL}$ measured under $H_z$ are similar. In this case, the spin current precesses in the $x$-$y$ plane under a small $H_z$. $R_{CSI}^{S_y}$ and $R_{CSI}^{S_x}$ are symmetric and antisymmetric against $H_z$. Their signs reverse after switching the initial direction of $M$. Therefore, $R_{CSI}^{S_y}$ and $R_{CSI}^{S_x}$ can be separated by symmetrizing and antisymmetrizing $R_{NL}^{diff}(H_z)$ when $H_z$ is scanned. Although $R_{CSI}^{S_z}$ vs. $H_z$ should have an $S$-shaped curve, similar to $R_{CSI}^{S_x}$ measured vs. $H_x$, a linear background exists in $R_{NL}$ because of the large ordinary Hall effect in Gr. Hence, it is difficult to determine $R_{CSI}^{S_z}$ using an $H_z$-dependent measurement.

First-principles calculations

We performed first-principles calculations based on density functional theory (DFT) [55] as implemented in Quantum Espresso package (QE) [56] [57]. The exchange-correlation functional was treated within the generalized gradient approximation of Perdew-Burke-Ernzerhof (PBE) [58]. The kinetic energy cutoff of electronic wavefunctions and charge density were chosen to be 80 and 640 Ry, respectively. To mimic the 2D layered structure, we included a large vacuum region in periodic cells, and used Grimme-D3 van der Waals correction [59] where necessary. The spin-orbit coupling was included in the electronic structure calculations. The k-grid meshes were chosen to be 15×15×9, 5×5×9, and 4×4×1 for bulk normal 1T-TaS$_2$, bulk CCDW 1T-TaS$_2$, and graphene/1T-TaS$_2$ heterostructures, respectively.

The intrinsic spin Hall conductivity $\sigma_{\alpha\beta}^{\gamma}$ was calculated by the Kubo-Greenwood formula as [60]

$$\sigma_{\alpha\beta}^{\gamma}(\varepsilon) = \frac{\hbar}{\Omega_C N_k} \sum_{\bm{k}} \sum_{n} f_{n\bm{k}} \sum_{m \neq n} \frac{2\text{Im}[\langle n\bm{k}|\hat{j}_\alpha^\gamma|m\bm{k}\rangle\langle m\bm{k}|-e\hat{v}_\beta|n\bm{k}\rangle]}{(\varepsilon_{n\bm{k}} - \varepsilon_{m\bm{k}})^2 + \eta^2},$$

where $\Omega_C$ and $N_k$ indicate the cell volume and the number of k-points used for k-space sampling, $f_{n\bm{k}}$ indicates Fermi-Dirac distribution, $\hat{j}_\alpha^\gamma$ is the spin current operator defined as $\hat{j}_\alpha^\gamma = \frac{1}{2}\{\hat{s}^\gamma, \hat{v}_\alpha\}$ and $\hat{s}^\gamma$ and $\hat{v}_{\alpha,\beta}$ are spin and velocity operators, respectively. $\eta$ is an adjustable smearing parameter. The numerical calculations were performed by Wannier90 package [61] [62] [63] [64]. For wannierization, *d*, *p*, and *p_z* orbital projections were used for Ta, S, and C atoms, and additional *sp²* orbital projections were alternatively added to the C atoms. To obtain the converged results, we used a fine k-mesh grid of 300×300×180 for normal 1T-TaS$_2$, 80×80×160 for CCDW 1T-TaS$_2$, and 600×600×1 for Gr/1T-TaS$_2$ heterostructures. The corresponding broadening constant of $\eta$ was chosen to be 10 meV. In the Gr/1T-TaS$_2$ heterostructures, we used an effective thickness of 9.46 Å when we evaluate their $\sigma_{\alpha\beta}^{\gamma}$ in units of $((\hbar/e)\,S/cm)$ (presented in Table 1).

**Acknowledgements**

We thank F. de Juan and H. Ochoa for the fruitful discussions. We acknowledge funding by the "Valleytronics" Intel Science Technology Center, the Spanish MCIN/AEI and by ERDF "A way of making Europe" (Project No. PID2021-122511OB-I00 and "Maria de Maeztu" Units of Excellent Programme No. CEX2020-001038-M), the European Union H2020 under the Marie Sklodowska-Curie Actions (Project Nos. 0766025-QuESTech and 955671-SPEAR), and Diputación de Gipuzkoa (Project No. 2021-CIEN-000037-01). S.L. and T.L. are partially supported by National Science Foundation through the University of Minnesota MRSEC under Award Number DMR-2011401. Z.C. and J.I.-A. acknowledge postdoctoral fellowship support from "Juan de la Cierva" Programme by the Spanish MCIN/AEI (grants No. FJC2021-047257-I and FJC2018-038688-I, respectively). S.L. is also supported by Basic Science Research Program through the National Research Foundation of Korea (NRF) funded by the Ministry of Education (NRF-2021R1A6A3A14038837). N.O. thanks the Spanish MCIN/AEI for support from a Ph.D. fellowship (Grant No. BES-2017-07963). B. M.-G. thanks support from "Ramón y Cajal" Programme by the Spanish MCIN/AEI (grant no. RYC2021-034836-I).

**Author contributions**

Z.C. and F.C. conceived this study. Z.C. fabricated and characterized the devices, performed the electrical measurements, and analyzed the data with the help of H.Y., E.D., C.K.S, J.I.-A., F.H., and N.O.. S.L. and T.L. performed the first-principles calculations. All authors contributed to the discussion on the results. Z.C., S.L., and F.C. wrote the manuscript with input from all authors.

**Competing interests**

The authors declare no competing interests.

Supplementary Information for

# Control of charge-spin interconversion in van der Waals heterostructures with chiral charge density waves

Zhendong Chi[1,*], Seungjun Lee[2,*], Haozhe Yang[1,*], Eoin Dolan[1], C. K. Safeer[1,3], Josep Ingla-Aynés[1], Franz Herling[1], Nerea Ontoso[1], Beatriz Martín-García[1,4], Marco Gobbi[4,5], Tony Low[2,6], Luis E. Hueso[1,4], Fèlix Casanova[1,4,†]

[1]CIC nanoGUNE BRTA, 20018 Donostia-San Sebastián, Basque Country, Spain

[2]Department of Electrical and Computer Engineering, University of Minnesota, Minneapolis, Minnesota 55455, USA

[3]Department of Physics, Clarendon Laboratory, University of Oxford, OX13PU, Oxford, United Kingdom

[4]IKERBASQUE, Basque Foundation for Science, 48009 Bilbao, Basque Country, Spain

[5]Centro de Física de Materiales (CSIC-EHU/UPV) and Materials Physics Center (MPC), 20018 Donostia-San Sebastián, Basque Country, Spain

[6]Department of Physics, University of Minnesota, Minneapolis, Minnesota 55455, USA

[*]These authors equally contributed to this work.

[†]E-mail: f.casanova@nanogune.eu

**Note 1. Atomic force microscopy characterization of 1T-TaS$_2$**

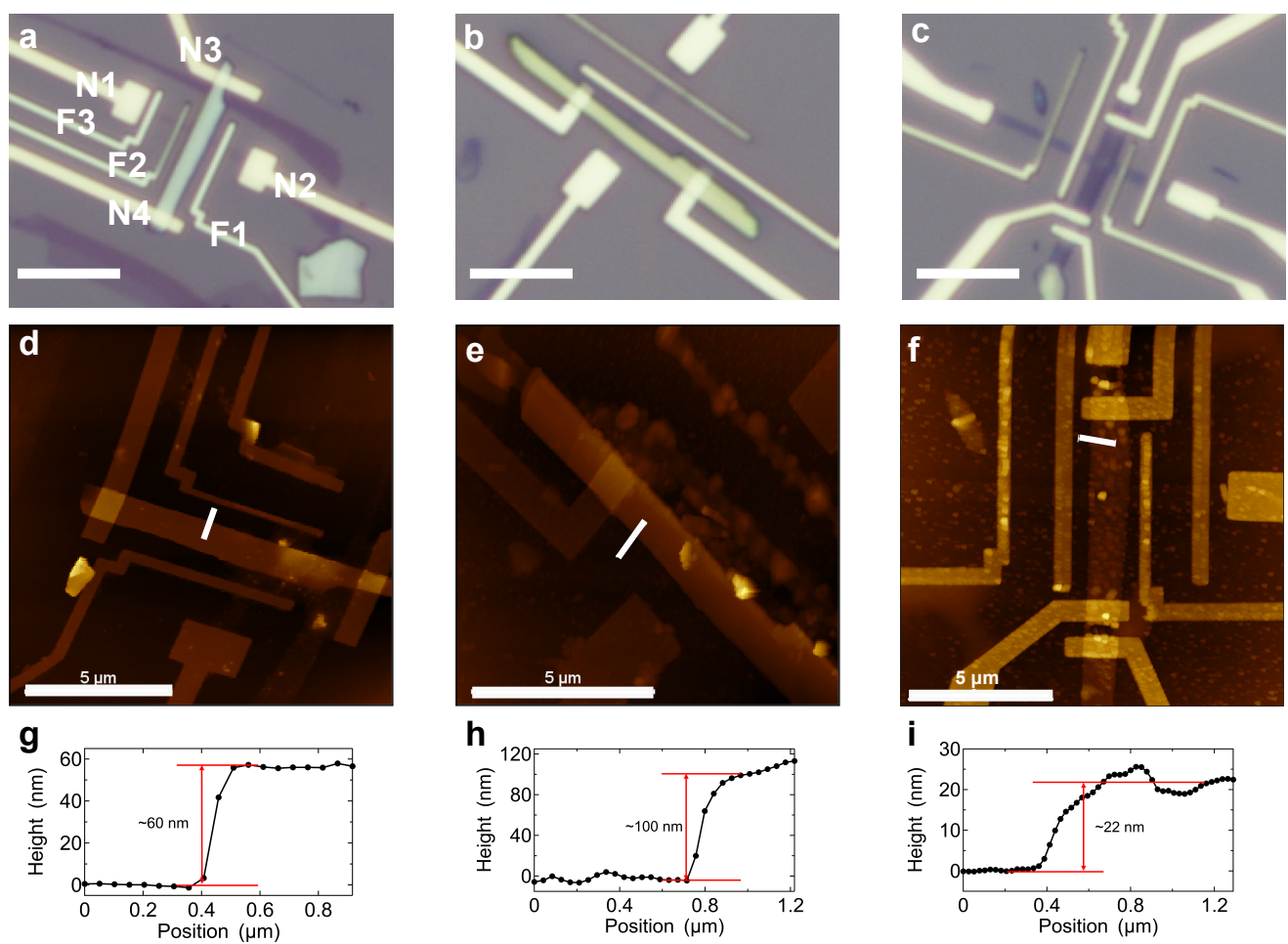

**Figure S1.** a-c, Optical images of Devices 1 (a), 2 (b), and 3 (c). Scale bar: 5 μm. The ferromagnetic (FM) electrodes (F1 to F3 on Gr) and non-magnetic (NM) electrodes (N1 and N2 on Gr, N3 and N4 on 1T-TaS$_2$) in Device 1 are labelled. d-f, Atomic force microscopy (AFM) images of Devices 1 (d), 2 (e), and 3 (f). g-i, The height profiles along the white lines drawn in d-f. The thicknesses of the 1T-TaS$_2$ flakes in the corresponding devices are tagged, as 60 nm (Device 1), 100 nm (Device 2), and 22 nm (Device 3).

## Note 2. Electrical transport of 1T-TaS$_2$ in Devices 2 and 3

The temperature dependence of the resistance of 1T-TaS$_2$ flake in the three devices is illustrated in Fig.1b in the main text (Device 1) and Fig. S2 (Devices 2 and 3). The data were collected using the two-points measurement method, e.g., between electrodes N3 and N4 in Device 1 (schematically displayed in Fig. S2c). Three resistance levels with two temperature hysteresis loops are observed below 360 K in Devices 1 and 2, which correspond to the ICCDW, NCCDW, CCDW phases and their phase transitions. The NCCDW-CCDW transition in Device 1 was stimulated by applying an in-plane high voltage to 1T-TaS$_2$ at 120 K [1]. In Device 3, only the ICCDW and NCCDW phases are observed in the 22-nm-thick 1T-TaS$_2$ flake. These results are in good agreement with previous studies, which have shown that the high-resistive CCDW phase disappears in thin (<40 nm) 1T-TaS$_2$ flakes [1] [2] [3] [4].

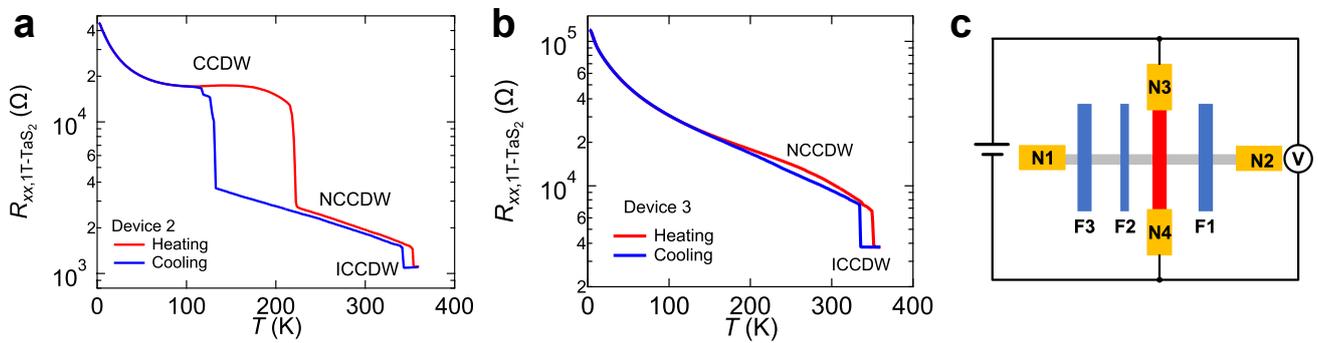

**Figure S2.** Temperature dependence of the longitudinal resistance of 1T-TaS$_2$ flakes ($R_{xx,1T-TaS2}$) in Devices 2 (a) and 3 (b). The heating and cooling processes are illustrated by the red and blue lines, respectively. The different CDW phases in the hysteresis loops are labeled. c, the sketch of the measurement geometry in Device 1.

**Note 3. Raman spectroscopy characterization on the graphene in Devices 1 and 2**

The number of graphene layers in Devices 1 and 2 is confirmed by using Raman spectroscopy (Alpha 300R Confocal Raman WITec microscope) with a 532 nm laser as excitation source, a diffraction grating of 600 1/nm, and 100× objective lens. Figure S3a illustrates the Raman spectra of the graphene in Devices 1 (red) and 2 (blue) in the range from 1400 to 2900 cm$^{-1}$. The spectra exhibit clear G and 2D graphene characteristic peaks. The Raman spectra zooming in the 2D peak range in Devices 1 and 2 are shown in Figs. S3b and S3c, respectively. The graphene 2D peak in Device 1 can be fitted well using four Lorentzians, indicating that it is a bilayer [5]. The graphene 2D peak in Device 2 can be fitted well using a single Lorentzian, suggesting that in this case it is a monolayer.

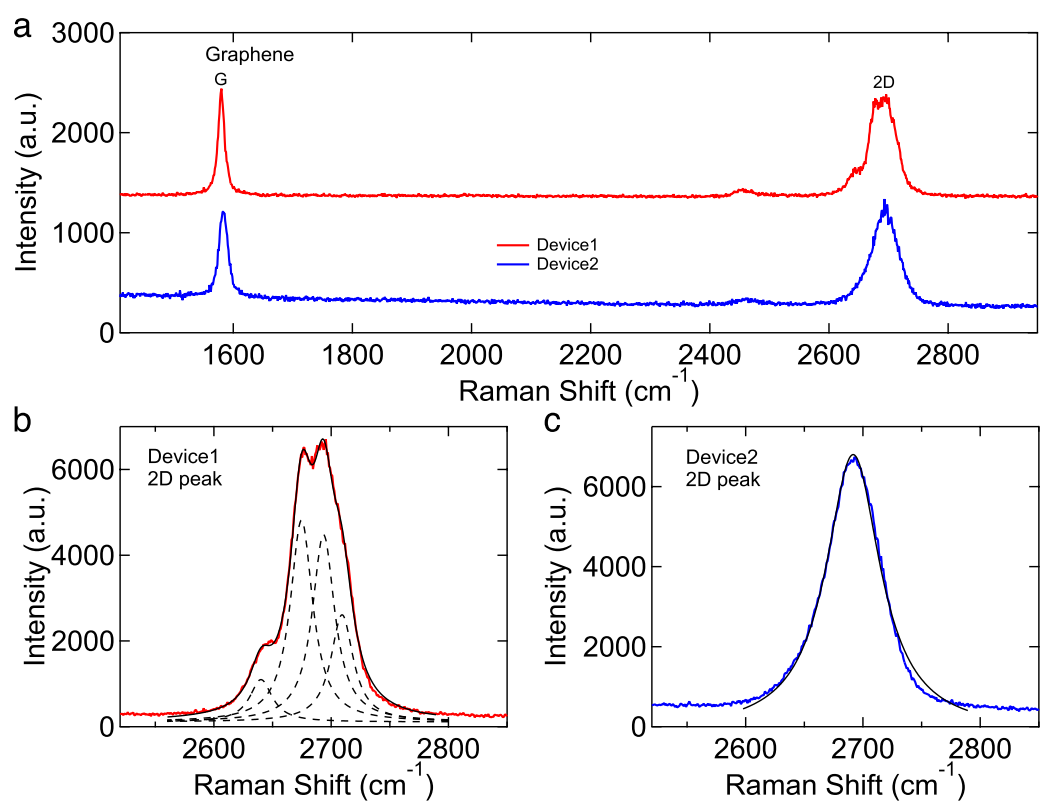

**Figure S3.** Raman spectra of the graphene in Devices 1 and 2. a, Raman spectra of the graphene in Devices 1 (red) and 2 (blue) showing the G and 2D peaks. b, c, Detailed Raman spectra in the 2D peak range in Devices 1 (b) and 2 (c). The black dashed curves in b represent the four Lorentzians used to fit the 2D peak, and the solid line is the sum of the four Lorentzians. The black solid line in c is the single Lorentzian fitting of the 2D peak of the graphene in Device 2.

## Note 4. Gate-voltage dependent transport in Device 1

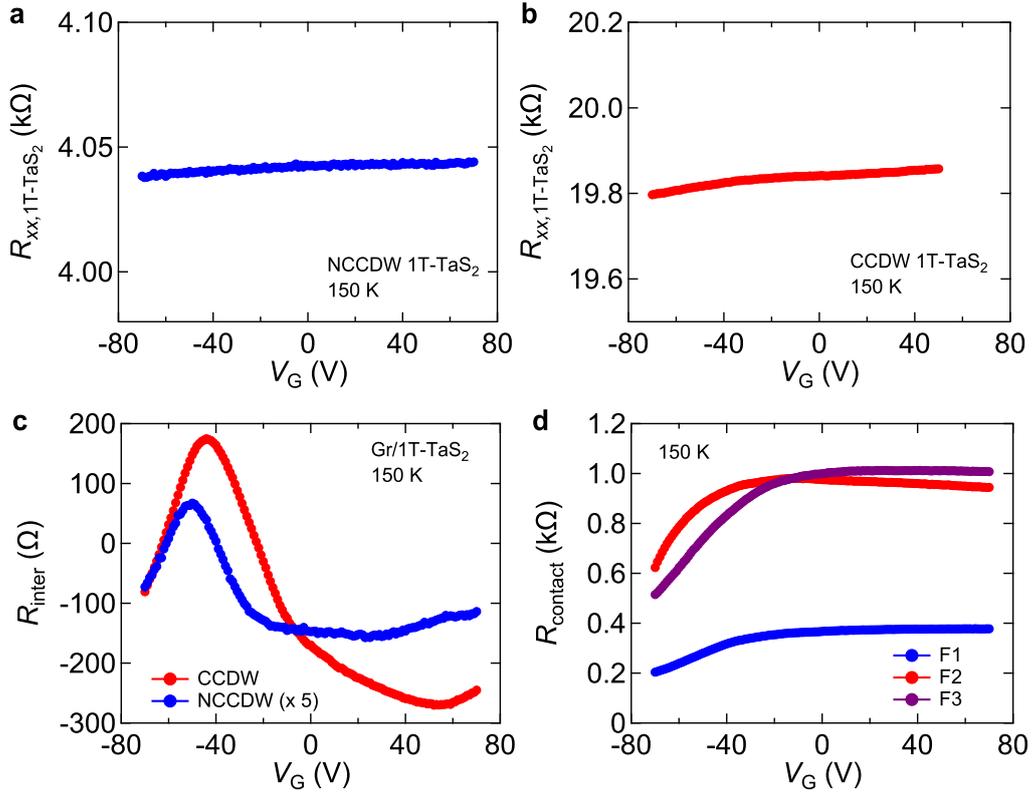

**Figure S4.** Gate voltage ($V_G$) dependence of transport properties of graphene/1T-TaS$_2$ heterostructure in Device 1. a,b, $V_G$-dependent longitudinal resistance of 1T-TaS$_2$ ($R_{xx,1T-TaS_2}$) in the NCCDW (a) and CCDW (b) phases. Two-point measurements carried out between N3 and N4 electrodes (see labels in Fig. S1a). c, $V_G$-dependent graphene/1T-TaS$_2$ interface resistance ($R_{inter}$). Four-point measurements were carried by adding a dc current between N4 and N2 electrodes and detecting the voltage between N1 and N3 electrodes (See labels in Fig. S1a). Red and blue curves represent 1T-TaS$_2$ in the CCDW and NCCDW phases, respectively. Blue curve is displayed by multiplying a factor of 5. d, $V_G$-dependence of the contact resistance ($R_{contact}$) of the three ferromagnetic electrodes F1 (blue), F2 (red), and F3 (purple). Three-point measurements were carried out by applying a dc current between N1 and F1 (F2, F3) electrodes and detecting the voltage between N2 and F1 (F2, F3) electrodes (See labels in Fig. S1a). $V_G$-dependent longitudinal resistance of Gr/1T-TaS$_2$ heterostructure ($R_{xx,cross}$) has been displayed in Fig. 4c in the main text. Four-point measurements were carried out by applying a dc current between electrodes N1 and N2 and detecting the voltage between F2 and F1 electrodes (see labels in Fig. S1a). $R_{xx,cross}$ is lower when 1T-TaS$_2$ is in the CCDW phase compared to that in the NCCDW phase, which agrees with a recent report *[6]*. All measurements were carried out at 150 K.

**Note 5. Spin transport characterization**

The spin injection efficiency from the FM electrodes to Gr is characterized by using the lateral spin valve (LSV) defined by two FM electrodes on Gr [7] [8] [9]. We measure the non-local resistance $R_{NL}$ for pristine Gr and Gr/1T-TaS$_2$ heterostructures using two FM electrodes without and with a 1T-TaS$_2$ flake placed between them, respectively (e.g., F2 and F3, F1 and F2 in Device 1). $R_{NL}$ of the LSV with pristine Gr in Device 1 measured at 100 K is plotted as a function of $H_y$ in Fig. S5a, showing a clear difference between the parallel (↑↑) and antiparallel (↑↓) configurations of the two FM electrodes. The non-local spin valve signal is defined as $\Delta R_{LSV} = R_{NL}^{\uparrow\uparrow} - R_{NL}^{\uparrow\downarrow}$, and Fig. S5b shows that $\Delta R_{LSV}$ of pristine Gr in Device 1 changes weakly with temperature.

$R_{NL}$ of the LSV across the Gr/1T-TaS$_2$ heterostructure in Device 1 measured at 150 K and without gate voltage is plotted as a function of $H_y$ in Fig. S5d. $R_{NL}$ measured when 1T-TaS$_2$ is in the NCCDW and CCDW phases are represented by the blue and red curves, respectively. We find that $\Delta R_{LSV}$ is almost zero in the NCCDW 1T-TaS$_2$ phase, while $R_{NL}$ shows a clear difference for ↑↑ and ↑↓ configurations in the CCDW 1T-TaS$_2$ phase, due to different spin absorption between these two phases. $\Delta R_{LSV}$ in the Gr/CCDW 1T-TaS$_2$ heterostructure is measured as a function of $V_G$ at 150 K and has been displayed in Fig. 4d in the main text.

Hanle precession is also measured as a function of $H_x$ for pristine Gr (Fig. S5c) and Gr/CCDW 1T-TaS$_2$ heterostructure (Fig. S5e) in Device 1. However, the oscillation of $R_{NL}$ due to precession (both in ↑↑ and ↑↓ configurations) is partially overlapped with magnetization saturation, making an accurate fitting of spin transport parameters difficult. This is caused by the fact that the two FM electrodes used in the measurement are too close (~2 $\mu m$).

We note that both non-local LSV and Hanle precession measurements substantiate the magnetic properties of the FM electrodes and verify the spin current injection characteristics. The mechanism of non-local LSV mirrors the giant magnetoresistance (GMR) effect induced by the spin current injected from the FM electrodes, leading to the two-step behavior of $R_{NL}$ corresponding to the alignment of the two FM electrodes, as shown in Figs. S5a and S5d. In Hanle precession measurements, the spins diffusing in the graphene channel from the FM electrode undergo a precession under $H_x$. This precession alters the *y*-axis projection of the polarization of the spin current, which in turn influences the voltage signal detected by the FM electrode due to the formation of an angle with its magnetization (***M***). Consequently, a decrease (increase) in the signal as a function of $H_x$ correlates with the initial parallel (antiparallel) alignment of the two FM electrodes, as shown in Figs. S5c and S5e, confirming that the signal stems from the spin current injection.

Moreover, the ***M*** properties of the FM electrodes can be determined by the Hanle precession non-local resistance $R_{NL}$ results due to the ***M*** pulling induced by $H_x$, which follows the Stoner-Wohlfarth model [10]. With $H_x$ perpendicular to the easy axis of the FM electrodes, the x-axis component of ***M*** ($M_x$) is proportional to $H_x/H_x^{sat}$, where $H_x^{sat}$ is the saturation magnetic field of ***M*** along the x-axis. The averaged $R_{NL}$ signal with parallel ($R_{NL}^P$) and antiparallel ($R_{NL}^{AP}$) initial alignments removes the precession component while preserving the magnetization properties, read as:

$$R_{avg} = \frac{R_{NL}^P + R_{NL}^{AP}}{2} = R_0 \sin^2 \theta_M. \tag{S1}$$

Here, $R_0$ is the magnetic signal when $H_x = 0$ and $\theta_M$ is the angle between ***M*** and the y-axis. Consequently, $\sin \theta_M$ can be written as:

$$\sin\theta_M = \text{sign}(H_x)\sqrt{\frac{R_{avg}-R_{avg}^{min}}{R_{avg}^{max}-R_{avg}^{min}}}, \tag{S2}$$

where $\text{sign}(H_x)$ is the sign of the applied magnetic field. We plot $\sin\theta_M$ determined from Hanle precession measurements on pristine graphene at 100 K as a function of $H_x$ in Fig. S5f. The result is in good agreement with the Stoner-Wohlfarth model, indicating a saturation of magnetization along the x-axis at $H_x \sim 2$ kOe. This curve has the same $H_x$-dependence as the $R_{CSI}^{S_x}$ component signal shown in Fig. 2i of the main text, evidencing that such CSI component is due to the $S_x$-component spin current injected from the FM electrode.

The characterization of the spin injection efficiency of Devices 2 and 3 is presented in Fig. S6. Figures S6a and S6b illustrate $R_{NL}$ of a lateral spin valve with pristine Gr measured at 300 K for Device 2 and Device 3, respectively. The temperature dependence of $\Delta R_{LSV}$ in Device 3 is shown in Fig. S6c. Notably, $\Delta R_{LSV}$ in these two devices are significantly larger than that observed in Device 1 (Figs. S5a and S5b), indicating a larger spin injection efficiency. $\Delta R_{LSV}$ of pristine Gr in Device 3 measured at 50 K as a function of $V_G$ is plotted in Fig. S6d. $\Delta R_{LSV}$ exhibits the opposite trend as a function of $V_G$ than the resistance of Gr, which is consistent with the behavior observed in Device 1 (Figs. 4c and 4d in the main text).

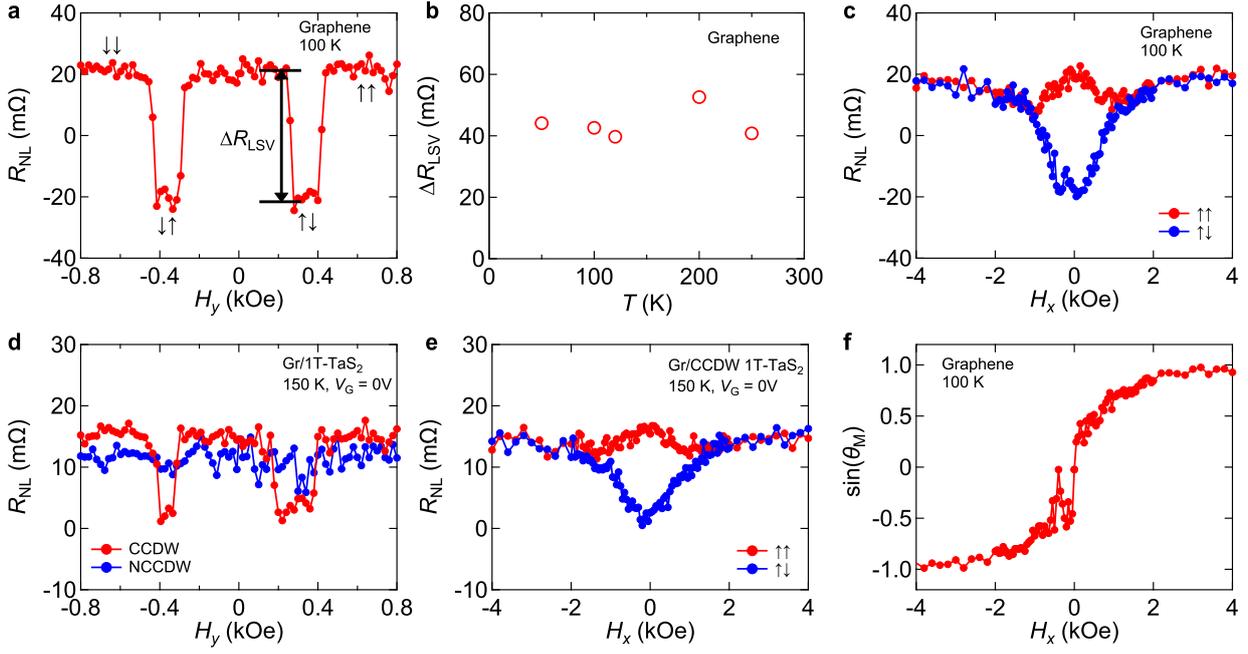

**Figure S5.** Spin transport characterization on Device 1. **a**, Non-local resistance ($R_{NL}$) of the LSV with pristine Gr measured as a function of $H_y$ at 100 K. ↑ and ↓ indicate the magnetization directions of the two FM electrodes used in the non-local spin valve measurement. The spin signal is defined as the difference between the parallel and antiparallel states and labeled as $\Delta R_{LSV}$. **b**, Temperature ($T$) dependence of $\Delta R_{LSV}$ in the LSV with pristine Gr. **c**, Hanle precession of the LSV with pristine Gr measured as a function of $H_x$ at 100 K. ↑↑ (↑↓) represent the parallel (antiparallel) configuration of the two FM electrodes in the LSV. **d**, $R_{NL}$ of the LSV with Gr/1T-TaS$_2$ heterostructure measured under $H_y$ at 150 K. The red and blue lines represent 1T-TaS$_2$ in CCDW and NCCDW phases, respectively. **e**, Hanle precession of the lateral spin valve with Gr/CCDW 1T-TaS$_2$ heterostructure measured under $H_x$ at 150 K. **f**, $\sin\theta_M$, where $\theta_M$ is the rotated angle of the magnetization of FM electrode with respect to the easy axis (y-axis), which is obtained by using Eq. S2.

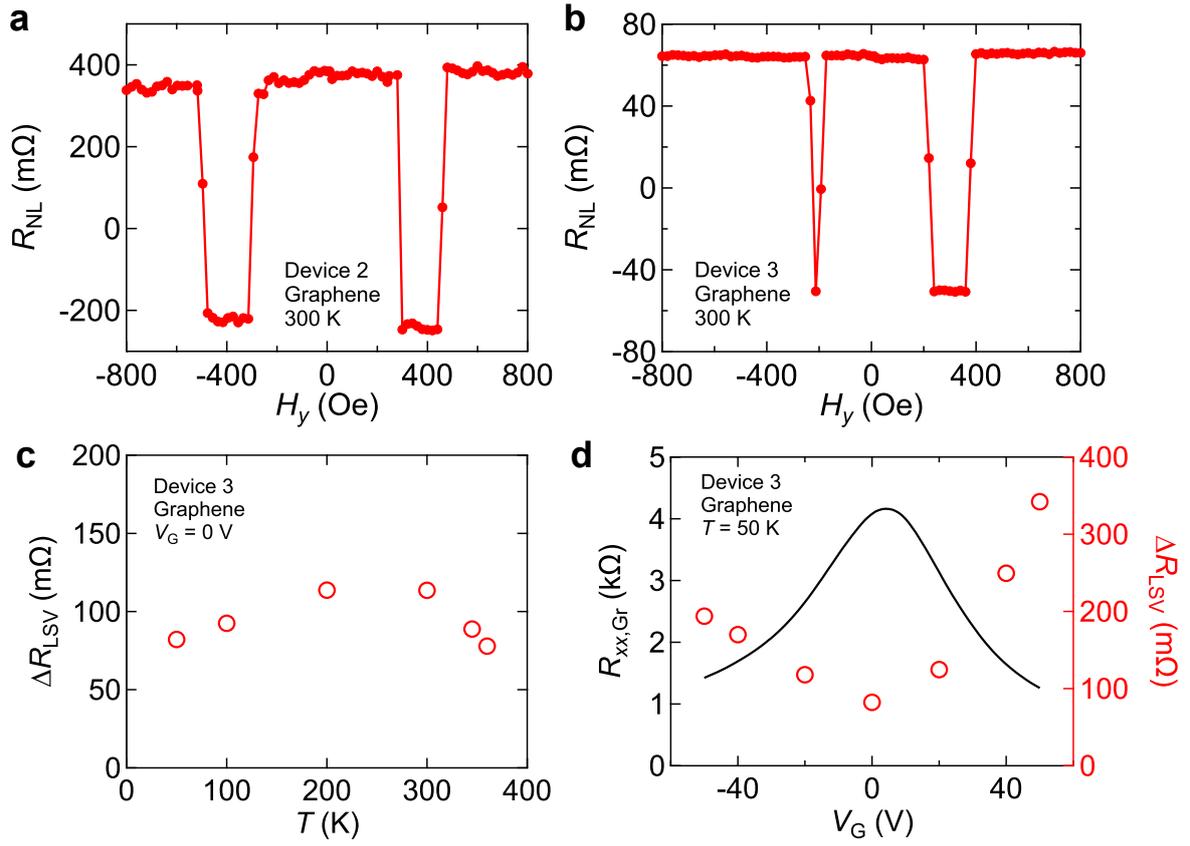

**Figure S6.** Spin transport characterization on Devices 2 and 3. a, b, Non-local resistance $R_{NL}$ of the lateral spin valve with pristine Gr as a function of $H_y$ at 300 K in Devices 2 (a) and 3 (b). The spin signal is defined as the difference between the parallel and antiparallel states and labeled as $\Delta R_{LSV}$. c, Temperature dependence of $\Delta R_{LSV}$ in the lateral spin valve with pristine Gr at $V_G = 0$ V in Device 3. d, $V_G$ dependence of the resistance ($R_{xx,Gr}$, black curve) and $\Delta R_{LSV}$ (red circles) of pristine Gr at 50 K in Device 3.

## Note 6. Raw data of the CSI characterization on Devices 1 & 2

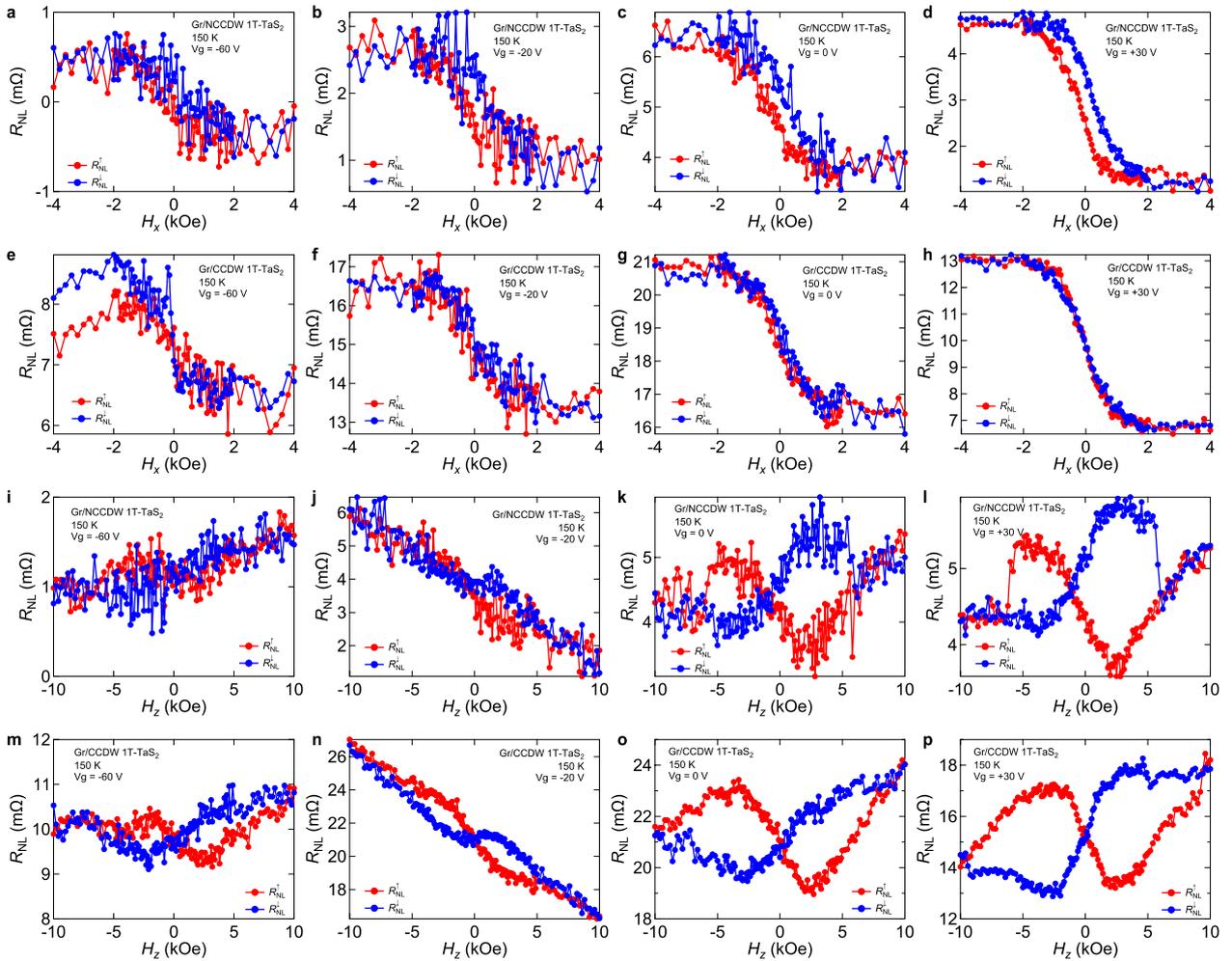

**Figure S7.** $R_{NL}$ of CSI raw data collected in Device 1 with $V_G$=−60, −20, 0, and +30 V at 150 K. a-d, $H_x$-dependent $R_{NL}$ in Gr/NCCDW 1T-TaS$_2$. e-h, $H_x$-dependent $R_{NL}$ in Gr/CCDW 1T-TaS$_2$. i-l, $H_z$-dependent $R_{NL}$ in Gr/NCCDW 1T-TaS$_2$. m-p, $H_z$-dependent $R_{NL}$ in Gr/CCDW 1T-TaS$_2$.

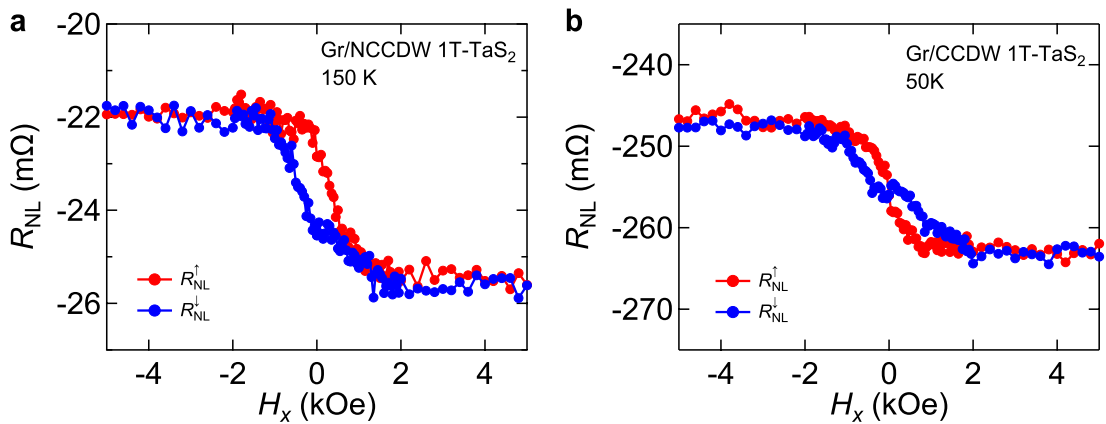

**Figure S8.** $R_{NL}$ raw data measured in Device 2. a, $H_x$-dependent $R_{NL}$ in Gr/NCCDW 1T-TaS$_2$ measured at 150 K and $V_G$= 0 V. b, $H_x$-dependent $R_{NL}$ in Gr/CCDW 1T-TaS$_2$ measured at 50 K and $V_G$= 0 V.

**Note 7.** $H_y$ dependent CSI hysteresis loops in Devices 1 & 2

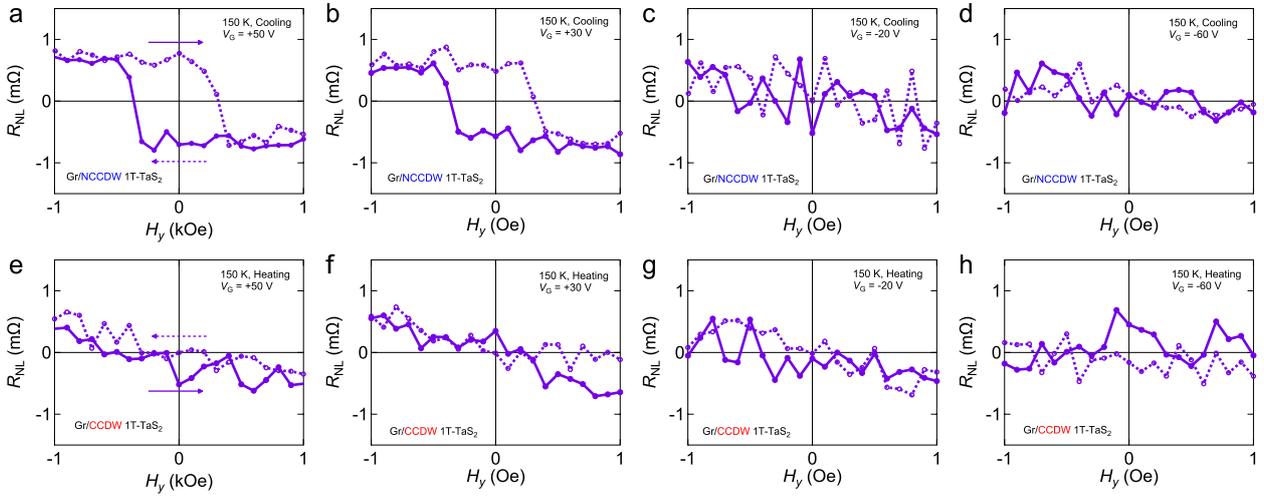

**Figure S9.** Non-local CSI resistance ($R_{NL}$) of Device 1 measured as a function of $H_y$ at the NCCDW (**a-d**) and CCDW (**e-h**) phases of 1T-TaS$_2$ with different back gate voltages ($V_G$). $V_G$ equals +50 (a,e), +30 (b,f), -20 (c,g), and -60 (d,h) V, respectively. The arrows represent the scanning directions of $H_y$. All the data were obtained at 150 K. A constant baseline has been subtracted from the lines in all the panels.

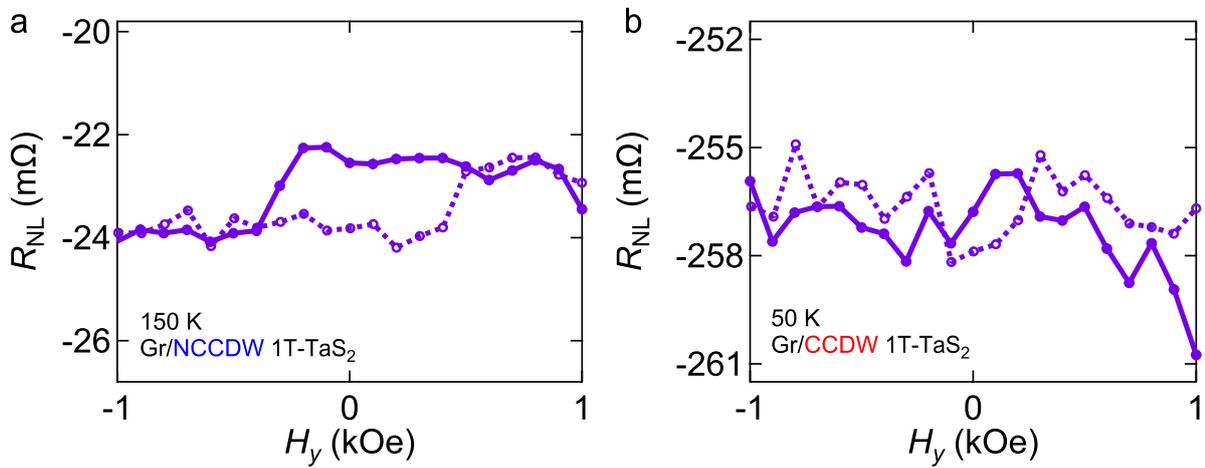

**Figure S10.** $R_{NL}$ of Device 2 measured as a function of $H_y$ at the NCCDW (a) and CCDW (b) phases of 1T-TaS$_2$. The measuring temperature is 150 K (a) and 50 K (b), respectively. No back gate voltage was added during the measurement.

**Note 8. CSI characterization on Device 3**

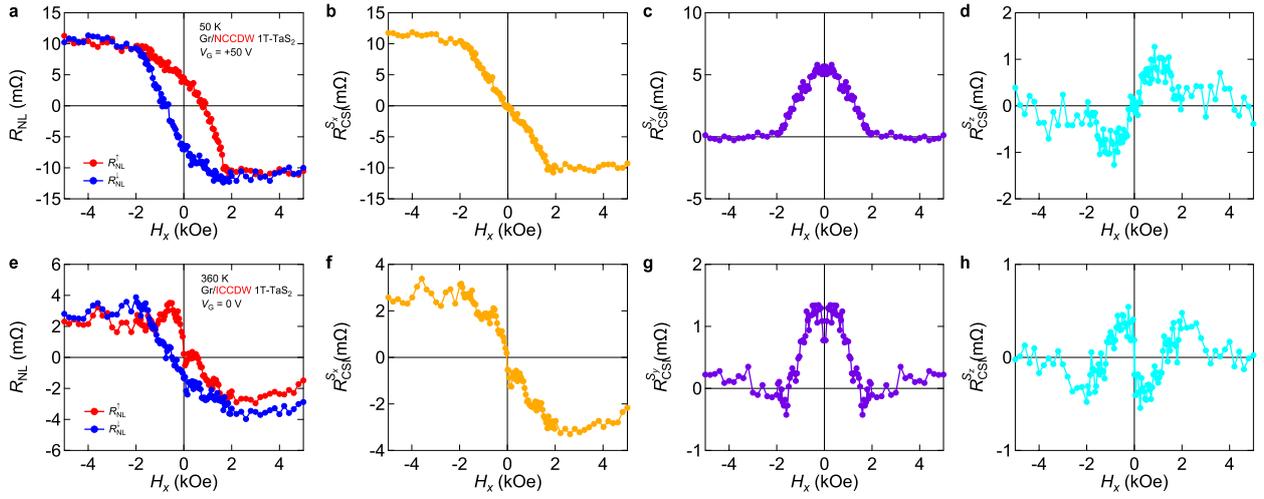

**Figure S11.** CSI characterization on Device 3. a-d, $H_x$-dependent $R_{NL}$ (a), $R_{CSI}^{S_x}$ (b), $R_{CSI}^{S_y}$ (c), and $R_{CSI}^{S_z}$ (d) in Gr/NCCDW 1T-TaS$_2$ heterostructure measured with $V_G = +50$ V at 50 K. e-h, $H_x$-dependent $R_{NL}$ (e), $R_{CSI}^{S_x}$ (f), $R_{CSI}^{S_y}$ (g), and $R_{CSI}^{S_z}$ (h) in Gr/ICCDW 1T-TaS$_2$ heterostructure measured with $V_G = 0$ V at 360 K.

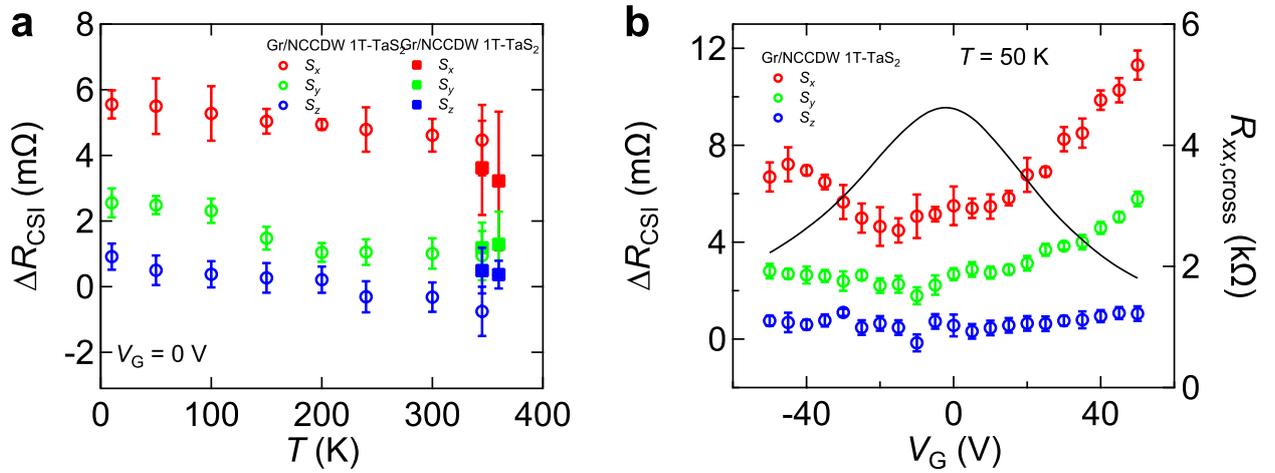

**Figure S12.** a, Temperature dependence of $\Delta R_{CSI}$ induced by $S_x$ (red), $S_y$ (green), and $S_z$ (blue) measured at $V_G = 0$ V. b, Gate voltage dependence of $\Delta R_{CSI}$ induced by $S_x$ (red), $S_y$ (green), and $S_z$ (blue) measured at $T = 50$ K. The black line is the resistance of Gr/1T-TaS$_2$ heterostructure ($R_{xx,cross}$).

**Note 9. First-principles calculations details**

The equilibrium lattice constants of normal and CCDW states of 1T-TaS$_2$ were calculated to be a = 3.356 Å, c = 5.792 Å, and a = 12.130 Å, c = 6.057 Å, respectively. For the bulk CCDW phase, we only considered the AA stacking configuration. We used a 5×5 supercell structure of graphene to construct a Gr/CCDW 1T-TaS$_2$ heterostructure as shown in Figure S12. The lattice mismatch between Gr and CCDW 1T-TaS$_2$ is about 1%. We also used the same size of the supercell for Gr/metallic 1T-TaS$_2$ heterostructure. Here, the twist-angle between graphene and 1T-TaS$_2$ is 13.9°.

Initially, we aim to understand the role of the interlayer interaction in the Gr/1T-TaS$_2$ heterostructure with respect to the CCDW phase. Figures S13a and S13b show their band structures and spin Berry curvature (SBC) along the high symmetry line. Here, the colors represent the conventional SBC components ($\Omega_{xy}^z$). In both cases, Gr is hole doped indicating charge transfer from Gr to 1T-TaS$_2$. Such a strong interaction leads to an SBC hotspot which is clearly shown in the inset of Figure S13a. Interestingly, in the CCDW phase, SBC hotspots are only observed in the Dirac band (near K points) not in the CDW band. However, in the normal state, not only Dirac bands but also 1T-TaS$_2$ bands contribute to SBC hotspots due to its broken symmetry and enhanced perpendicular electric field in 1T-TaS$_2$ assisted by Gr.

The spin Hall conductivity (SHC) tensor was calculated by Kubo formula as described in the Methods section. As a first step, we constructed a Wannier Hamiltonian by using Wannier90 code [70] [71] [13]. As shown in Figure S14, our Wannier band structures show excellent agreement with band structures calculated by first-principles calculations. Then, we evaluated the SHC as a function of energy (Figure S14) and constructed the SHC tensors by using the SHC values at the Fermi level (Figure S15).

Although the SHC components are very sensitive to the in-plane rotation, it is very difficult to determine the exact crystallographic orientation of both graphene and 1T-TaS$_2$ in the experiments. To thoroughly investigate all possibilities, we evaluated crystal-orientation-dependent SHC, as shown in Figure S15, obtained through $\sigma_{[a'_1 a'_2 a'_3]_{ij}}^{k} = \sum_{lmn} D_{il} D_{jm} D_{kn} \sigma_{[a_1 a_2 a_3]_{lm}}^{n}$, where $D$ is a rotation matrix and $a_1 a_2 a_3$ and $a'_1 a'_2 a'_3$ indicate crystal orientation before and after rotation [14] [15]. Then, the final SHC components represented in Table 1 of the main text were chosen from Figure S14 giving a maximum value of each component.

The Fermi surface and spin textures displayed in Figure 5 of the main text and Figure S18 were calculated by Wannier90 code. In order to visualize chirality-dependent spin textures more clearly, we also used the Vienna ab initio simulation package (VASP) [15] and calculated spin textures of graphene's Dirac bands slightly above Fermi level, as shown in Figure S19. The same crystal structures as in Figure 5 of the main text were used.

In addition, for a more thorough verification, we calculated the spin textures of another Gr/CCDW 1T-TaS$_2$ heterostructure with a twist-angle of 5.2° between graphene and 1T-TaS$_2$, as shown in Figure S20. It is worth noting that, in this example, two opposite-handedness structures no longer have the mirror relation. Nevertheless, their radial spin textures are still almost cancelled out, suggesting that our discussion on chirality-dependent spin textures is robust in the general case.

**Note 10. Discussion on the spin-orbit proximity in the graphene/1T-TaS$_2$ heterostructure**

A comparison of the strength of the spin-orbit proximity and CSI between our graphene/1T-TaS$_2$ heterostructures and other graphene/semiconducting TMDC heterostructure systems might be helpful. However, we note that current redistribution between graphene and 1T-TaS$_2$ dramatically enhances the number of potential mechanisms for CSI, including bulk spin Hall effect (SHE) in 1T-TaS$_2$, interfacial SHE, and Rashba-Edelstein effect (REE). This complexity makes it challenging to directly compare our system to other graphene/semiconducting TMDC heterostructures.

Furthermore, the relationship between the strength of the spin-orbit proximity, i.e., valley-Zeeman coupling, and the interfacial SHE, is not directly linked. Accurately estimating the value of interfacial SHE and determining the efficiency of REE in the heterostructure system necessitates complex simulations that account for both valley-Zeeman and Rashba couplings.

To evaluate the strength of spin-orbit proximity in graphene, we calculated the spin-resolved band structures of both Gr/normal 1T-TaS$_2$ and Gr/CCDW 1T-TaS$_2$ heterostructures, as shown in Figure S21. Near the Dirac point, the calculated band structure was fitted to the continuum Hamiltonian [16] [17]:

$$H(\kappa \mathbf{K} + \mathbf{k}) = \hbar v_F(\kappa \sigma_x k_x + \sigma_y k_y) + \Delta \sigma_z + \lambda_R e^{-\frac{is_z\phi}{2}}(\kappa \sigma_x s_y - \sigma_y s_x)e^{\frac{is_z\phi}{2}} + \kappa \lambda_{VZ}\sigma_0 s_z, \quad (S3)$$

where $v_F$ is the Fermi velocity, $\Delta$ is the staggered potential due to sublattice asymmetry, $\kappa = \pm 1$ is the valley index (for valley $\pm$K), and $\lambda_R$ and $\lambda_{VZ}$ are the Rashba and valley-Zeeman coupling terms, respectively. The $\sigma_i$ ($s_i$), with $i = 0, x, y, z$, indicate the orbital (spin) matrices and $\phi$ is the Rashba angle parameter accounting for the non-orthogonal spin-momentum locking in Rashba bands [17]. For the Gr/normal 1T-TaS$_2$ heterostructure, the proximitized band was nicely fitted to Eq. S3 and the fitting parameters were calculated to be $\Delta = 0.10$ meV, $\lambda_R = 0.67$ meV, and $\lambda_{VZ} = 0.59$ meV, respectively. However, we realized that the proximitized band structure of Gr/CCDW 1T-TaS$_2$ heterostructure was not able to be fitted to Eq. S3 because the symmetry of CCDW 1T-TaS$_2$ is completely different from its pristine counterpart.

Although we can compare the $\lambda_{VZ}$ and $\lambda_R$ values with other systems, such as in the case of a graphene/WS$_2$ heterostructure where $\lambda_{VZ} = 1.12$ meV and $\lambda_R = 0.36$ meV [18], we note that the potential CSI mechanisms are intricate in our system given the conductive nature of 1T-TaS$_2$. Therefore, it may not be appropriate to directly compare our experimental findings with those systems including semiconducting TMDC, for instance, the graphene/WS$_2$ heterostructure [19].

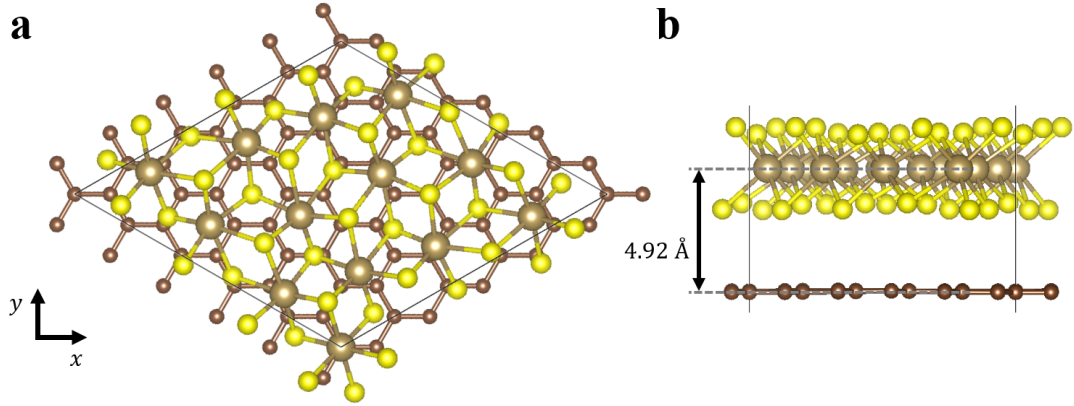

**Figure S13.** Crystal structure of Gr/CCDW 1T-TaS$_2$ heterostructure. **a**, Top and **b**, side view of heterobilayer. Gold, yellow, and brown spheres indicate Ta, S, and C atoms, respectively.

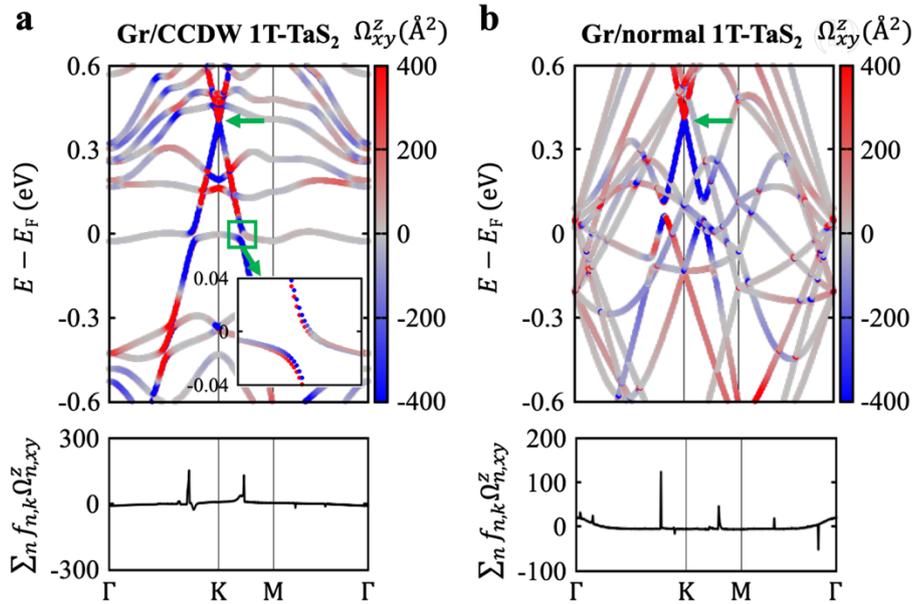

**Figure S14.** First-principles calculations for spin Berry curvature. **a, b,** Spin Berry curvature ($\Omega_{xy}^z$) projected electronic structures and integrated $\Omega_{xy}^z$ for Gr/1T-TaS$_2$ heterostructures with and without CCDW phase. In each panel, green arrows indicate Dirac point of Gr. In a, the inset shows spin Berry curvature hotspots originating from interaction between Dirac and CCDW bands.

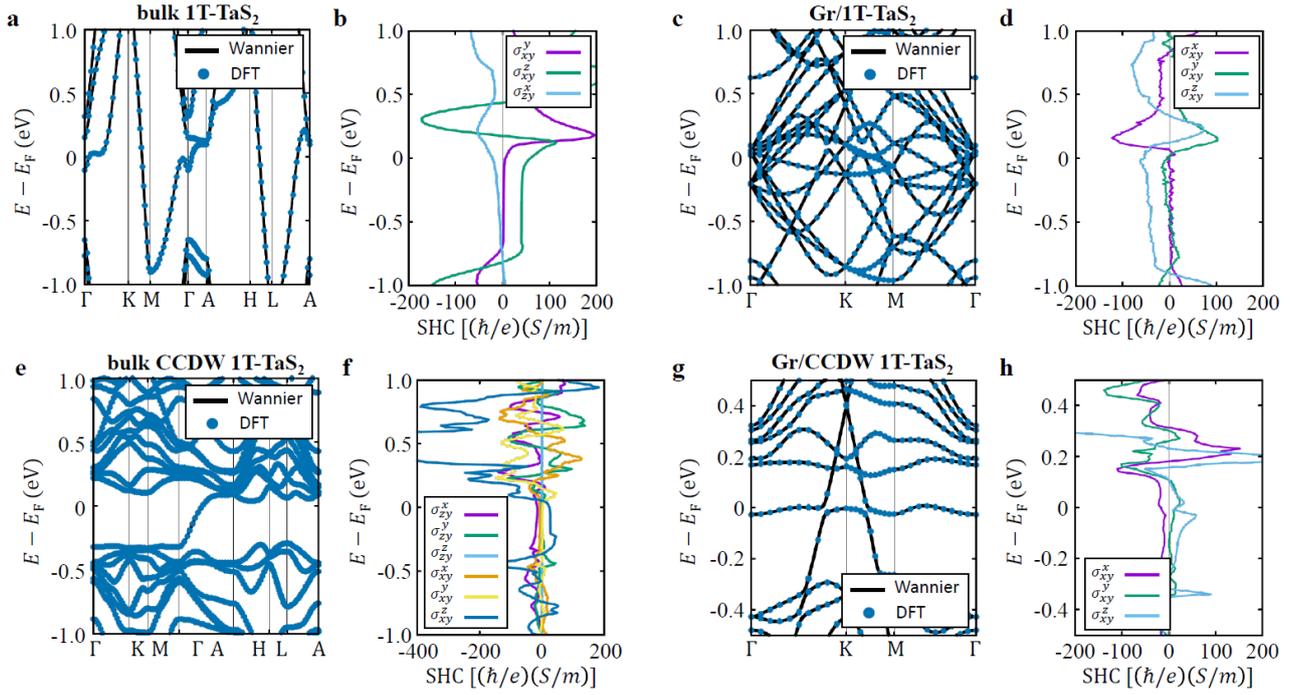

**Figure S15.** Electronic structure and selected SHC components of **a, b,** bulk normal 1T-TaS$_2$, **c, d,** Gr/normal 1T-TaS$_2$ heterostructure, **e, f,** bulk CCDW 1T-TaS$_2$, and **g, h,** Gr/CCDW 1T-TaS$_2$ heterostructure. In **a, c, e, g,** solid black line shows the band structure obtained by Wannier90 showing excellent agreement blue dots obtained from DFT calculations.

**a** bulk 1T-TaS$_2$

$$\sigma_{ij}^x = \begin{pmatrix} -6.5 & 0 & 0 \\ 0 & 6.5 & 20.6 \\ 0 & -22.0 & 0 \end{pmatrix}$$

$$\sigma_{ij}^y = \begin{pmatrix} 0 & 6.5 & -20.6 \\ 6.5 & 0 & 0 \\ 22.0 & 0 & 0 \end{pmatrix}$$

$$\sigma_{ij}^z = \begin{pmatrix} 0 & 59.6 & 0 \\ -59.6 & 0 & 0 \\ 0 & 0 & 0 \end{pmatrix}$$

**b** Gr/1T-TaS$_2$

$$\sigma_{ij}^x = \begin{pmatrix} -0.26 & -6.08 \\ 0.16 & <0 \end{pmatrix}$$

$$\sigma_{ij}^y = \begin{pmatrix} -5.35 & 1.28 \\ 1.97 & 0.60 \end{pmatrix}$$

$$\sigma_{ij}^z = \begin{pmatrix} -9.7 & -52.8 \\ 53.7 & -11.58 \end{pmatrix}$$

**c** bulk CCDW 1T-TaS$_2$

$$\sigma_{ij}^x = \begin{pmatrix} -2.9 & -8.4 & 4.1 \\ 1.8 & 1.8 & 21.5 \\ -3.3 & -15.7 & <0.1 \end{pmatrix}$$

$$\sigma_{ij}^y = \begin{pmatrix} -3.2 & 1.8 & -22.0 \\ 2.2 & 2.8 & 3.2 \\ 15.6 & -2.9 & <0.1 \end{pmatrix}$$

$$\sigma_{ij}^z = \begin{pmatrix} 0.4 & 23.0 & 0.3 \\ -24.2 & -0.3 & <0.1 \\ 0.4 & <0.1 & -0.1 \end{pmatrix}$$

**d** Gr/CCDW 1T-TaS$_2$

$$\sigma_{ij}^x = \begin{pmatrix} -13.07 & -16.64 \\ -4.62 & 10.01 \end{pmatrix}$$

$$\sigma_{ij}^y = \begin{pmatrix} -14.54 & 16.35 \\ 5.96 & 10.17 \end{pmatrix}$$

$$\sigma_{ij}^z = \begin{pmatrix} 1.83 & 14.70 \\ -22.17 & 2.23 \end{pmatrix}$$

**Figure S16.** Calculated SHC tensor at $E = E_F$ for **a,** normal 1T-TaS$_2$, **b,** Gr/normal 1T-TaS$_2$ heterostructure, **c,** CCDW 1T-TaS$_2$, **d,** and Gr/CCDW 1T-TaS$_2$ heterostructure in units of $(\hbar/e)\, S/cm$. "$<0.1$" indicates that the absolute value of that component is smaller than 0.1 $(\hbar/e)\, S/cm$.

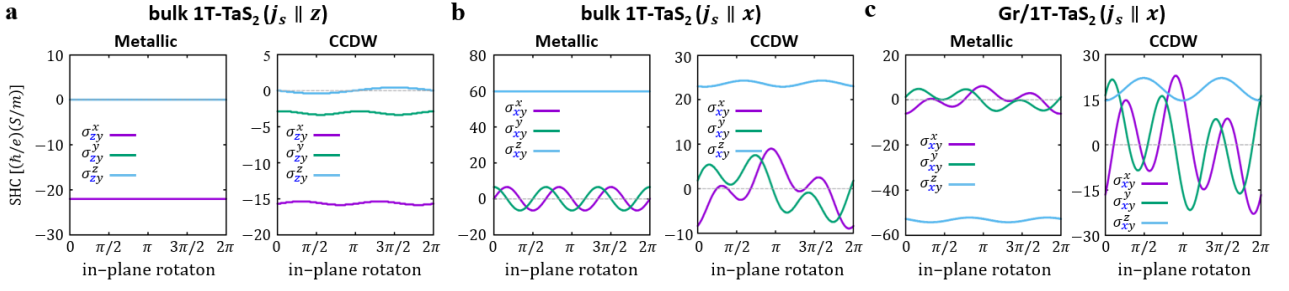

**Figure S17.** Selected SHC components corresponding to our experimental observation as function of in-plane rotation angle. Here, direction of *x* at the zero-rotation angle indicates a zigzag direction for the normal 1T-TaS$_2$, as illustrated in Figure S12 for CCDW 1T-TaS$_2$.

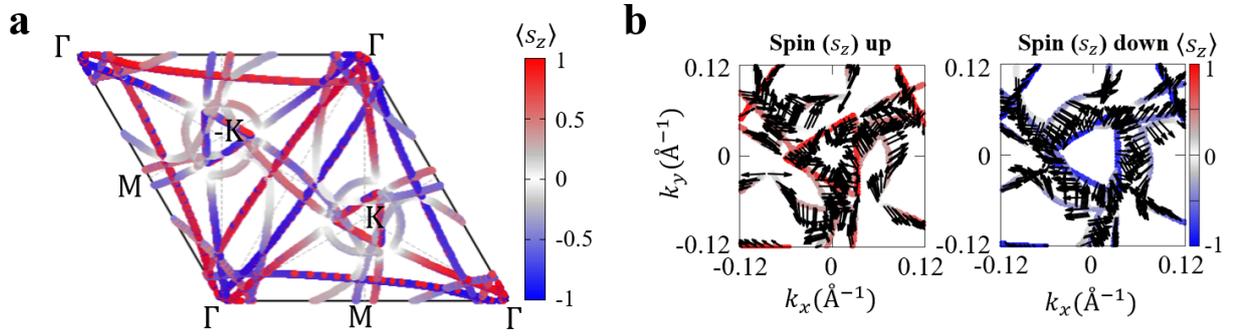

**Figure S18.** Spin-resolved Fermi surface of **a,** Gr/normal 1T-TaS$_2$ heterostructure and **b,** its in-plane spin distribution near K point.

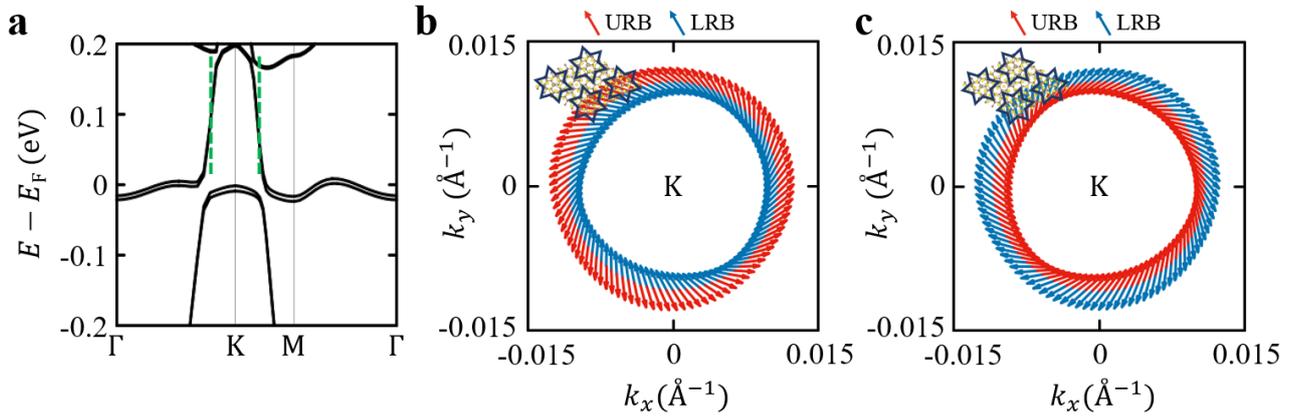

**Figure S19. a,** Electronic structures of Gr/CCDW 1T-TaS$_2$ calculated by VASP. **b, c,** In-plane spin distributions of graphene's Dirac bands near K point for two opposite-handedness structures, respectively. In **b, c,** red and blue arrows indicate upper and lower Rashba bands (URB and LRB) split from the Dirac bands of graphene due to spin-orbit proximity. For a clear comparison, we plot spin textures along the circle near K point, and the radius of the circle is represented as a green dashed line in **a**.

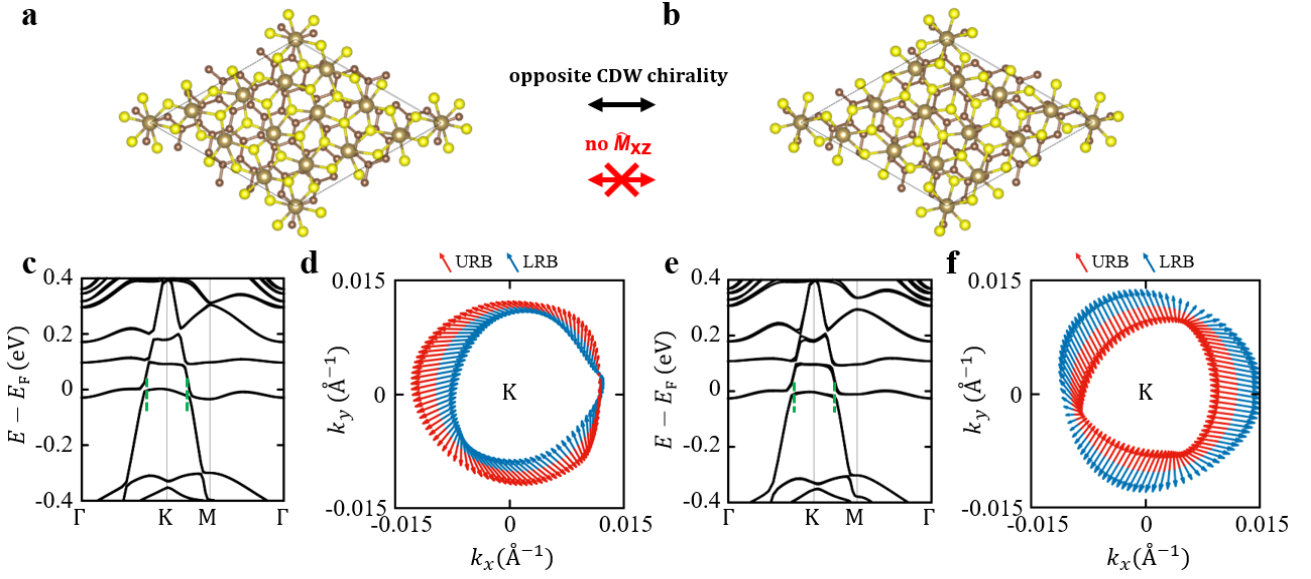

**Figure S20. a, b,** Another two possible chiral states of Gr/CCDW 1T-TaS$_2$ heterostructures with opposite handedness. Here, the twist angle between graphene and 1T-TaS$_2$ is 5.2°. Handedness-dependent **c, e,** electronic structures and **d, f,** in-plane spin textures of proximitized Dirac bands near the K point. In **d** and **f,** red and blue arrows indicate the upper and lower Rashba bands (URB and LRB) split from graphene's Dirac bands due to spin-orbit proximity. For a clear comparison, we plot spin textures along the circle near K point, and the radius of the circle is represented as a green dashed line in **c** and **e**.

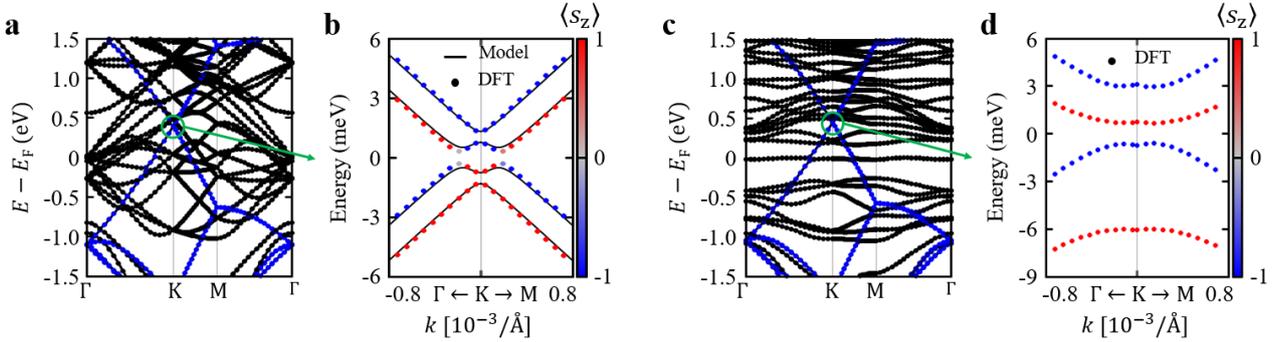

**Figure S21. a, c,** Electronic structures of Gr/normal 1T-TaS$_2$ and Gr/CCDW 1T-TaS$_2$ heterostructures, respectively. In **a** and **c,** blue color indicates electronic bands mainly originating from graphene and green circles represent Dirac points. **b, d,** Band structures near Dirac points. The color bar indicates the spin expectation value of its out-of-plane components. In **b,** the colored dots indicate energy eigenvalues obtained by our DFT calculation, which were fitted by the model defined in Eq. S3, plotted with black solid lines.